%% file: main.tex
\documentclass[twocolumn]{aastex631}
\usepackage{booktabs}
\usepackage{amssymb}
\usepackage{graphicx}
\usepackage{longtable}

\usepackage{hyperref}
\usepackage{verbatim}
\usepackage{tabularx}
\usepackage{xspace}	
\usepackage{xcolor}
\newcommand{\unit}[1]{\ensuremath{\mathrm{\,#1}}\xspace}
\newcommand{\feh}{\mathrm{[Fe/H]}}
\newcommand{\teff}{\ensuremath{T_\mathrm{eff}}\xspace}
\newcommand{\logg}{\ensuremath{\log\,g}\xspace}

\newcommand{\kms}{\unit{km\,s^{-1}}}

\shorttitle{Eri III \& DELVE 1: C-rich Clusters or Dwarfs?}
\shortauthors{Simon et al.}

\graphicspath{{./figures/}{}}

\begin{document}

\title{Eridanus III and DELVE 1: Carbon-rich Primordial Star Clusters or the Smallest Dwarf Galaxies?\footnote{This paper includes data gathered with the 6.5 meter Magellan Telescopes located at Las Campanas Observatory, Chile.}}

\author[0000-0002-4733-4994]{Joshua D.~Simon}
\affil{Observatories of the Carnegie Institution for Science, 813 Santa Barbara St., Pasadena, CA 91101, USA}

\author[0000-0002-9110-6163]{Ting~S.~Li}
\affiliation{Department of Astronomy \& Astrophysics, University of Toronto, 50 St. George Street, Toronto ON, M5S 3H4, Canada}
\affiliation{Dunlap Institute for Astronomy \& Astrophysics, University of Toronto, 50 St George Street, Toronto, ON M5S 3H4, Canada}

\author[0000-0002-4863-8842]{Alexander P. Ji}
\affiliation{Department of Astronomy \& Astrophysics, University of Chicago, 5640 S Ellis Avenue, Chicago, IL 60637, USA}
\affiliation{Kavli Institute for Cosmological Physics, University of Chicago, Chicago, IL 60637, USA}

\author[0000-0002-6021-8760]{Andrew~B.~Pace}
\affiliation{McWilliams Center for Cosmology, Carnegie Mellon University, 5000 Forbes Ave, Pittsburgh, PA 15213, USA}

\author[0000-0001-6154-8983]{Terese T.~Hansen}
\affil{Department of Astronomy, Stockholm University, AlbaNova
University Center, SE-106 91 Stockholm, Sweden}

\author[0000-0003-1697-7062]{William~Cerny}
\affil{Department of Astronomy, Yale University, New Haven, CT 06520, USA}

\author[0000-0002-9933-9551]{Ivanna~Escala}
\affiliation{Department of Astrophysical Sciences, Princeton University, 4 Ivy Lane, Princeton, NJ 08544, USA}
\affiliation{Observatories of the Carnegie Institution for Science, 813 Santa Barbara St., Pasadena, CA 91101, USA}

\author[0000-0003-2644-135X]{Sergey E. Koposov}
\affiliation{Institute for Astronomy, University of Edinburgh, Royal Observatory, Blackford Hill, Edinburgh EH9 3HJ, UK}
\affiliation{Institute of Astronomy, University of Cambridge, Madingley Road, Cambridge CB3 0HA, UK}
\affiliation{Kavli Institute for Cosmology, University of Cambridge, Madingley Road, Cambridge CB3 0HA, UK}

\author[0000-0001-8251-933X]{Alex~Drlica-Wagner}
\affiliation{Fermi National Accelerator Laboratory, P.O.\ Box 500, Batavia, IL 60510, USA}
\affiliation{Kavli Institute for Cosmological Physics, University of Chicago, Chicago, IL 60637, USA}
\affiliation{Department of Astronomy \& Astrophysics, University of Chicago, 5640 S Ellis Avenue, Chicago, IL 60637, USA}

\author[0000-0003-3519-4004]{Sidney~Mau}
\affiliation{Department of Physics, Stanford University, 382 Via Pueblo Mall, Stanford, CA 94305, USA}
\affiliation{Kavli Institute for Particle Astrophysics \& Cosmology, P.O. Box 2450, Stanford University, Stanford, CA 94305, USA}

\author[0000-0001-6196-5162]{Evan~N.~Kirby}
\affiliation{Department of Physics and Astronomy, University of Notre Dame, 225 Nieuwland Science Hall, Notre Dame, IN 46556, USA}

\begin{abstract}\label{sec:abstract}
We present spectroscopy of the ultra-faint Milky Way satellites Eridanus~III (Eri~III) and DELVE~1.  We identify eight member stars in each satellite and place non-constraining upper limits on their velocity and metallicity dispersions.  The brightest star in each object is very metal-poor, at $\feh = -3.1$ for Eri~III and $\feh = -2.8$ for DELVE~1.  Both of these stars exhibit large overabundances of carbon and very low abundances of the neutron-capture elements Ba and Sr, and we classify them as CEMP-no stars.  Because their metallicities are well below those of the Milky Way globular cluster population, and because no CEMP-no stars have been identified in globular clusters, these chemical abundances could suggest that Eri~III and DELVE~1 are dwarf galaxies.  On the other hand, the two systems have half-light radii of 8~pc and 6~pc, respectively, which is more compact than any known ultra-faint dwarfs.  We conclude that Eri~III and DELVE~1 are either the smallest dwarf galaxies yet discovered, or they are representatives of a new class of star clusters that underwent chemical evolution distinct from that of ordinary globular clusters.  In the latter scenario, such objects are likely the most primordial star clusters surviving today.  These possibilities can be distinguished by future measurements of carbon and/or iron abundances for larger samples of stars or improved stellar kinematics for the two systems.

\end{abstract}


\section{Introduction} \label{sec:intro}

For many decades, dwarf galaxies and globular clusters appeared to be clearly distinct classes of stellar systems, with no ambiguity about their nature and differences.  Globular clusters were once regarded as the quintessential simple stellar populations, with uniform chemical abundances for all of their stars, sizes on the order of $\sim1$~pc, and no dynamical evidence for dark matter from their stellar kinematics \citep[e.g.,][]{illingworth76,pl86}.  Subsequently, multiple chemically distinct populations of stars have been identified in many clusters, primarily affecting light elements \citep[e.g.,][]{cohen78,kraft79,peterson80,kraft92,gratton12}, although age differences within these clusters must be quite small.  Dwarf galaxies, contrastingly, have long been known to contain stars with a range of metallicities (and often ages as well), with radii $\gtrsim100$~pc, and stellar velocities that are best explained by the presence of a dark matter halo \citep[e.g.,][]{aaronson83,grebel97}.

In more recent years, this dichotomy has broken down with the discovery of ultra-faint dwarf galaxies and star clusters spanning the gap that previously existed between the size distributions of the classical dwarf spheroidals and globular clusters \citep[e.g.,][]{willman05a,willman05b,zucker06,belokurov07,belokurov09,koposov07}.  Classification of these objects has generally relied on the criterion proposed by \citet{ws12}: to be considered a galaxy, an object should have dynamical evidence of dark matter and/or exhibit chemistry that requires a deep enough gravitational potential to retain supernova ejecta.

\begin{figure}\centering
\includegraphics[width=0.52\textwidth]{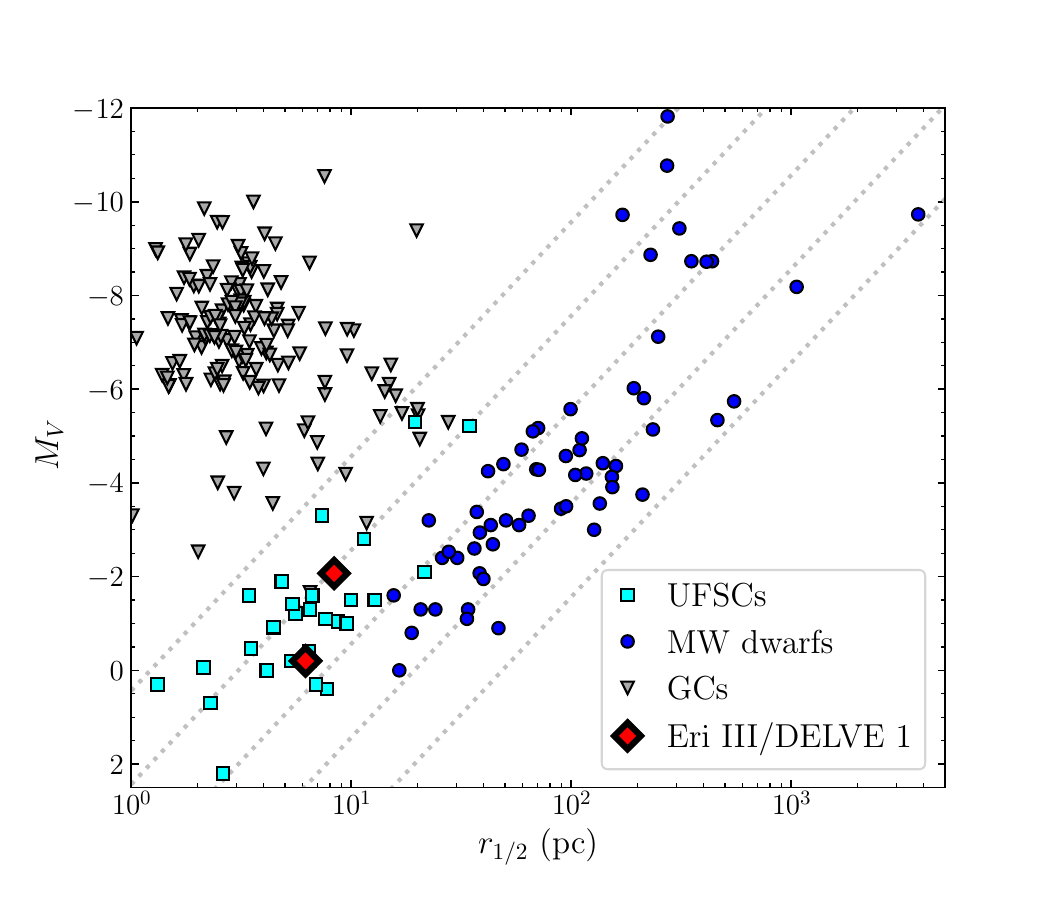}
\caption{Distribution of absolute magnitude as a function of half-light radius for stellar systems in the Milky Way halo.  Dwarf galaxies are displayed as blue circles, globular clusters listed in the \citet{harris10} compilation are plotted as gray triangles, and recently discovered ultra-faint star clusters are shown as cyan squares.  The subjects of this paper, Eri~III and DELVE~1, are represented by larger red diamonds, nominally located within the distribution of ultra-faint clusters.
\label{fig:mv_rhalf}}
\end{figure}

Considering the population of Milky Way satellites known now, the distributions of half-light radii for dwarf galaxies and globular clusters largely remain distinct at absolute magnitudes brighter than $M_{V} \sim -2$ ($L \sim 540$~L$_{\odot}$), with the dwarfs having $r_\mathrm{half} \gtrsim 20$~pc and clusters having $r_\mathrm{half} \lesssim 10$~pc \citep[e.g.,][]{belokurov07,McConnachie12,simon19}.  However, as shown in Fig.~\ref{fig:mv_rhalf}, there are some clusters with radii approaching or even exceeding 20~pc, generally at $M_{V} \lesssim -5$, such as AM~1, Eridanus, Pyxis, Palomar~3, Palomar~4, Palomar~14, Palomar~15, Crater, and Sagittarius~II \citep{munoz18,mp18,longeard20}.  On the other hand, several objects with similar sizes and fainter magnitudes, including Tucana~III, Grus~II, and Draco~II, are suspected to be dwarfs \citep{simon17,simon20,hansen17,hansen20,longeard18,fu23}, but conclusive evidence has been difficult to obtain.
At $M_{V} \gtrsim -2$, the dwarf galaxy sequence in the size-luminosity plane intersects the cluster locus, and classifications are currently highly uncertain (Fig.~\ref{fig:mv_rhalf}; also see \citealt{smith24}).  Furthermore, recent theoretical models suggest that the existence of dwarf galaxies with parameters in this range should be expected \citep[e.g.,][]{mk22,errani23}

The subjects of this paper, Eridanus~III ($M_{V} = -2.1 \pm 0.5$, $r_{{\rm half}} = 8.6^{+0.9}_{-0.8}$~pc; \citealt{conn18a}) and DELVE~1 ($M_{V} = -0.2^{+0.8}_{-0.6}$, $r_{{\rm half}} = 6.2^{+1.5}_{-1.1}$~pc; \citealt{mau20})\footnote{The original \citet{mau20} physical half-light radius for DELVE~1 was converted from the measured angular half-light radius for an assumed distance of 19~kpc.  As discussed in \S~\ref{sec:delve1}, we derived a revised distance of 22~kpc, so here we have updated the physical half-light radius to account for that difference.}, fall in this ambiguous region of parameter space.  Their sizes are small enough that it would be reasonable (or at least, consistent with the literature) to presume on that basis alone that they are faint clusters.  However, as we will demonstrate in the rest of this paper, the chemical properties of both systems are distinct from those of the known cluster population, resulting in some doubt as to their true nature.  In the following section, we present our spectroscopy of Eri~III and DELVE~1.  In Section~\ref{sec:results}, we describe our velocity and metallicity measurements for the IMACS data and evaluate which stars are members of Eri~III and DELVE~1.  We derive chemical abundances from high-resolution spectra in Section~\ref{sec:abundances}.  In Section~\ref{sec:interp}, we compare the properties of Eri~III and DELVE~1 to globular clusters and dwarf galaxies and discuss possible classifications of the two objects.  We summarize the paper and conclude in Section~\ref{sec:summary}.

\section{Observations and Data Reduction}
\label{sec:obs}

\subsection{Targets}

Eri~III was identified in the first year of data from the Dark Energy Survey (DES) by \citet{bechtol15} and \citet{koposov15}.  These initial studies noted its much smaller size compared to the other Milky Way satellites discovered in DES data, but did not draw any conclusions about its classification.  The reanalysis of deeper DES photometry by \citet{munoz18}, as well as new Gemini imaging by \citet{conn18a}, resulted in a somewhat smaller half-light radius, although the measurements from all four papers are consistent within the uncertainties.  DELVE~1 was discovered in early data from the DECam Local Volume Exploration Survey (DELVE) by \citet{mau20}, who suggested that it was a faint cluster and named it accordingly.  Neither satellite has thus far been the subject of any spectroscopic investigations, although \citet{fu23} derived photometric metallicities using narrow-band photometry for 13 stars in Eri~III and found evidence for a metallicity spread.  As part of a larger effort to characterize the faint star clusters that have been discovered in recent surveys (Cerny et al., in prep.), we obtained medium-resolution spectroscopy of Eri~III and DELVE~1 with the IMACS spectrograph \citep{dressler06,dressler11} on the Magellan/Baade telescope.  

\subsection{IMACS 1200-line Spectroscopy}
\label{sec:imacs1200}

We observed both satellites with the $f$/4 camera of IMACS, using the 1200~$\ell$/mm grating blazed at 9000~\AA\ to provide $R \approx 11,000$ spectra around the near-infrared Ca triplet (CaT) lines.  We observed Eri~III on 2020 February 2 (2.33~hours) and 2020 December 17 (1.0~hours), with a slit mask targeting the brightest red giant member candidate as well as two blue horizontal branch (BHB) stars, all of which were selected based on DES photometry and proper motions from the early data release 3 (EDR3) of Gaia \citep{lindegren21,gaiadr3vallenari}.

Similarly, we observed a DELVE~1 slit mask on 2021 June 18 (2.58~hours), 2021 June 27 (2.67~hours), 2022 June 20 (1.81~hours), and 2023 March 25--26 (4.5~hours).  The DELVE~1 target stars included two red giants and several subgiants and stars near the main sequence turnoff.  These stars were selected using astrometry and photometry from Gaia EDR3 and DECaLS DR9 \citep{dey19}; we did not use DELVE photometry because it is missing several likely member stars apparent in the Gaia catalog as a result of a 13th magnitude star located very close to DELVE~1.

The Eri~III and DELVE~1 observations are listed in Table~\ref{tab:imacs}.  Our observing and calibration procedures followed those described in previous papers \citep[e.g.,][]{simon17,simon20,li17,li18}.  As in related past work with IMACS, we reduced the data using a combination of the COSMOS package \citep{oemler17} and our modified version of the DEEP2 data reduction pipeline \citep{cooper12,newman13}.

\begin{deluxetable*}{c c c c c c c c}
\tablecolumns{9}
\centering
\tablewidth{0pt}
\tablecaption{\hypertarget{Table 1}{Summary of IMACS Spectroscopy}}\label{tab:imacs}
\tablehead{
\colhead{System} & \colhead{Date} & \colhead{MJD} & \colhead{R.A.}& \colhead{Decl.} & \colhead{$t_{\rm exp}$ } & \colhead{Seeing} &  \colhead{Grating}\\
{}&{}&{}&{(h:m:s)}&{(d:m:s)}&{(hours)}&{}&}
\startdata
Eri III & Feb 2, 2020 & 58882.1 & 02:22:53.00 & $-$52:19:04.92 & 2.33 & 0\farcs7 & 1200/32.7 \\
Eri III & Dec 17, 2020 & 59201.2 & 02:22:53.00 & $-$52:19:04.92 & 1.00 & 0\farcs5 & 1200/32.7 \\
Eri III & Sept 12-13, 2021 & 59471.0 & 02:23:01.00 & $-$52:14:00.00 & 6.93 & 0\farcs5"-1\farcs0 & 600/13.0 \\
\hline
DELVE 1 & Jun 18, 2021 & 59384.2 & 16:30:50.40 & $-$00:55:19.19 & 2.58 & 0\farcs8 & 1200/32.7 \\
DELVE 1 & Jun 27, 2021 & 59393.2 & 16:30:50.40 & $-$00:55:19.19 & 2.67 & 0\farcs7-1\farcs3 & 1200/32.7 \\
DELVE 1 & Jun 20, 2022 & 59751.1 & 16:30:50.40 & $-$00:55:19.19 & 1.81 & 0\farcs9-2\farcs0 & 1200/32.7 \\
DELVE 1 & Mar 25-26, 2023 & 60030.0 & 16:30:50.40 & $-$00:55:19.19 & 4.50 & 0\farcs5-0\farcs8  & 1200/32.7 \\
\enddata
\end{deluxetable*}

\subsection{IMACS 600-line Spectroscopy}

Following the observations described above, we observed Eri~III again with IMACS with a lower-resolution grating to obtain spectra of a larger sample of candidate member stars at fainter magnitudes ($g > 22$).  For these observations, we employed the 600~$\ell$/mm grating blazed at 7500~\AA\ to obtain $R \approx 5,000$ spectra covering the wavelength range 6000--9000~\AA.  We observed an Eri~III slit mask for 6.9~hours on the nights of 2021 September 12--13.  The data were obtained and reduced with the same procedures described in \S~\ref{sec:imacs1200}, using an updated line list appropriate for the broader wavelength range and lower spectral resolution.

\subsection{MIKE Spectroscopy}

After confirming that the brightest candidate star in each of Eri~III and DELVE~1 is indeed a member of the respective system with the IMACS spectra (see \S~\ref{sec:results}), we observed both stars at high spectral resolution with the MIKE spectrograph \citep{bernstein03} on the Magellan/Clay telescope to determine their chemical abundance patterns.  For the Eri~III observations, we used the 1\arcsec-wide slit, providing a spectral resolution of $R\approx 28,000$ in the blue spectrograph and $R \approx 22,000$ in the red spectrograph.  We observed the Eri~III star for 7.8~hours on 2020 October 19--20 and 2.4~hours on 2020 November 16.

We observed the substantially brighter DELVE~1 star with the 0\farcs7 slit to increase the spectral resolution to $R\approx40,000$ in the blue and $R\approx32,000$ in the red.  We obtained 1.5~hours of integration time on 2022 March 7.

The MIKE observations of both stars were reduced with a recent version of the pipeline originally introduced by \citet{kelson03}.  A summary of the data, including signal-to-noise ratio measurements, is given in Table~\ref{tab:mike}.

\begin{deluxetable*}{lccccccccccc}
\tablecolumns{12}
\tabletypesize{\footnotesize}
\tablecaption{\label{tab:mike}MIKE Targets and Observations}
\tablehead{Name & Gaia source\_id & R.A. & Decl. & $g_0$ & $r_0$ & Slit & $t_{\rm exp}$ & SNR & SNR  & $v_{\textrm{hel}}$ & MJD \\
{}&{}&{h:m:s}&{d:m:s}&(mag)&(mag)&{}&(hours)&(4500{\AA})&(6500{\AA})&{(\kms)}&{} }
\startdata
DELVE~1-S1 & 4359353258111283328 & 16:30:54.25 & $-$00:58:01.4 & 16.35 &  15.74 & 0\farcs7 & 1.5 & 34 & 59 & $-400.7 \pm 0.5$ & 59646.30 \\
Eri~III-S1 & 4745740262792352128 & 02:22:48.66 & $-$52:17:05.1 & 19.33 &  18.63 & 1\farcs0 & 10.2\tablenotemark{a} & 14\tablenotemark{a} & 39\tablenotemark{a} & \phs$\phn52.3 \pm 1.4$ & 59142.59\tablenotemark{b} \\
 &  &  &  & & & & & & & \phs$\phn53.5 \pm 1.5$ & 59170.17 
\enddata
\tablenotetext{a}{Combined value for coadded Eri~III-S1 spectrum from multiple nights.}
\tablenotetext{b}{This velocity measurement was made from a spectrum obtained over two consecutive nights; the MJD value reported here corresponds to the average over all exposures, which occurs halfway in between the two nights.}
\end{deluxetable*}

\section{Kinematics, Metallicities, and Membership from Medium-Resolution Spectroscopy}
\label{sec:results}

\subsection{Velocity Measurements}

Following the procedures established in previous studies \citep[e.g.,][]{simon17,li17}, we measured stellar velocities from the IMACS 1200-line data via $\chi^{2}$ fits to the template stars HD~122563 (for red giants) and HD~161817 (for BHB stars) over the wavelength range containing the Ca triplet lines (as well as the Paschen lines for the hotter BHB stars).  We corrected for mis-centering of the stars in their slits with a $\chi^{2}$ fit to the telluric A-band absorption using the B9 star HR~4781 as a template. 

We adopted similar procedures to make kinematic measurements with the IMACS 600-line data.  As for the 1200-line templates described by \citet{simon17}, we observed a set of template stars by orienting a long slit in the north-south direction and driving the star across the slit during an exposure to ensure that the star light uniformly filled the slit.  The minimum exposure time for the procedure was 4~minutes so that sufficient sky lines would be available to check the wavelength solution.  We again used HD~122563 as the red giant template, with HD~86986 as the BHB template, and the B5 star HR~1244 as the A-band template.

We do not currently have enough repeat measurements of stars with the 600-line grating to robustly determine the systematic velocity uncertainty of this spectrograph configuration.  However, based on the decrease in spectral resolution and the increase in the measured scatter of arc lamp lines around the best-fit wavelength solution (typically $\sim1$~\kms) relative to the 1200-line grating, we assume a systematic error floor of 3~\kms\ for 600-line data.

\subsection{Ca Triplet Metallicity Measurements}
\label{sec:cat}

We used the IMACS spectra with both gratings to compute metallicities from the equivalent widths (EWs) of the Ca triplet (CaT) lines for red giant stars, following the same procedures as in previous papers \citep[e.g.,][]{simon17,li17}.  We relied on the \citet{carrera13} CaT calibration to convert apparent magnitudes to absolute magnitudes, assuming distances of 91~kpc  for Eri~III \citep{conn18a} and 22~kpc for DELVE~1 (see Section \ref{sec:delve1}).  We measured EWs with a combined Gaussian+Lorentzian fit to each of the CaT lines.

DELVE~1 contains two RGB stars, for which we determine metallicities of $\feh = -2.84 \pm 0.12$ (brighter star) and $\feh = -2.40 \pm 0.26$ (fainter star).  Eri~III contains six RGB members, but for the five stars near the bottom of the giant branch, our spectra have S/N too low for reliable EW measurements.  We therefore stack these spectra to obtain a single metallicity measurement.  The brightest Eri~III star has a metallicity of $\feh = -3.20 \pm 0.17$.  The next brightest star on the RGB is near the S/N limit for reliable CaT EW measurements, but we estimate $\feh = -3.02 \pm 0.44$, compatible with the metallicity of the brightest member.  Although the remaining RGB members do not have sufficient S/N for individual EW measurements, after coadding the four spectra together, we find $\feh = -2.63 \pm 0.42$.  The slightly higher metallicity of the coadded spectrum could suggest a metallicity spread in Eri~III (as determined photometrically by \citealt{fu23}), but is also consistent at $\lesssim1.5\sigma$ with the metallicities of the brighter stars.

The IMACS velocity and metallicity measurements are listed in Table~\ref{tab:imacs_spec_table}.  Based on visual inspection of the fits, we consider the velocity measurements reliable down to $\textrm{S/N} = 4$ for Eri~III and $\textrm{S/N} = 3.8$ for DELVE~1.

\input{imacs_spec_table_membersonly}

\subsection{Membership Determinations and Systemic Properties}

\subsubsection{Eridanus III}
Based on an initial photometric and astrometric selection, three stars (one on the RGB and two on the BHB) appeared likely to be members of Eri~III.  From our 1200-line IMACS spectroscopy, we determined velocities for these three stars that are within 10~\kms\ of one another at a heliocentric velocity of $\sim50-60$~\kms, confirming their association with Eri~III.\footnote{The two BHB stars have very low S/N ratios of $\sim3$~pixel$^{-1}$, but because the Paschen lines provide signal over nearly all pixels between 8400~\AA\ and 8900~\AA, it is possible to measure velocities from quite low S/N spectra.}  We note that the brightest RGB star is somewhat redder than the appropriate isochrone.  Presuming that this star is an Eri~III member (which is strongly supported by the chemical abundance data described below), its red color would be most naturally explained by a high carbon abundance, which suppresses flux in the $g$ filter, resulting in a larger $g-r$ color (see, e.g., \citealt{hayes23}).

\begin{figure*}\centering
\includegraphics[width=0.24\textwidth]{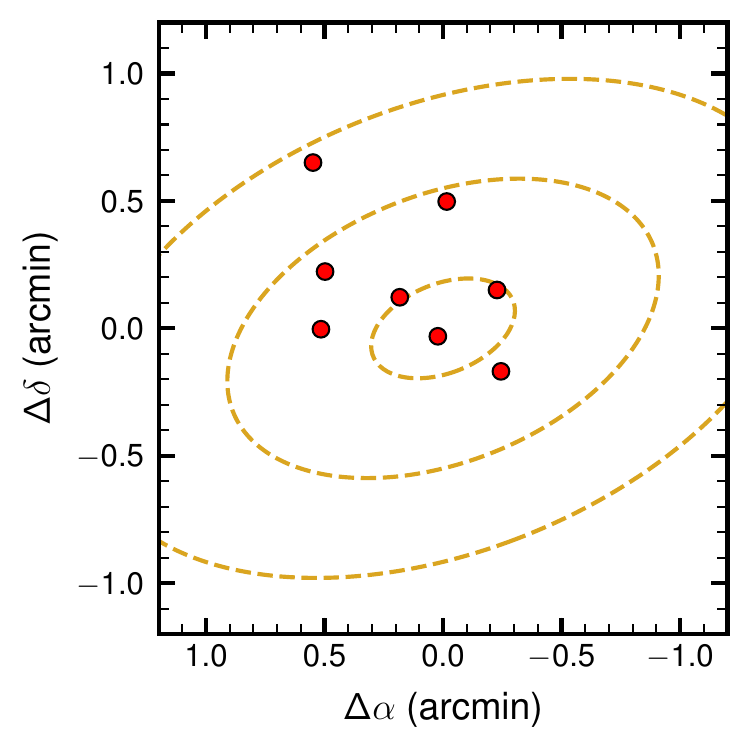}
\includegraphics[width=0.24\textwidth]{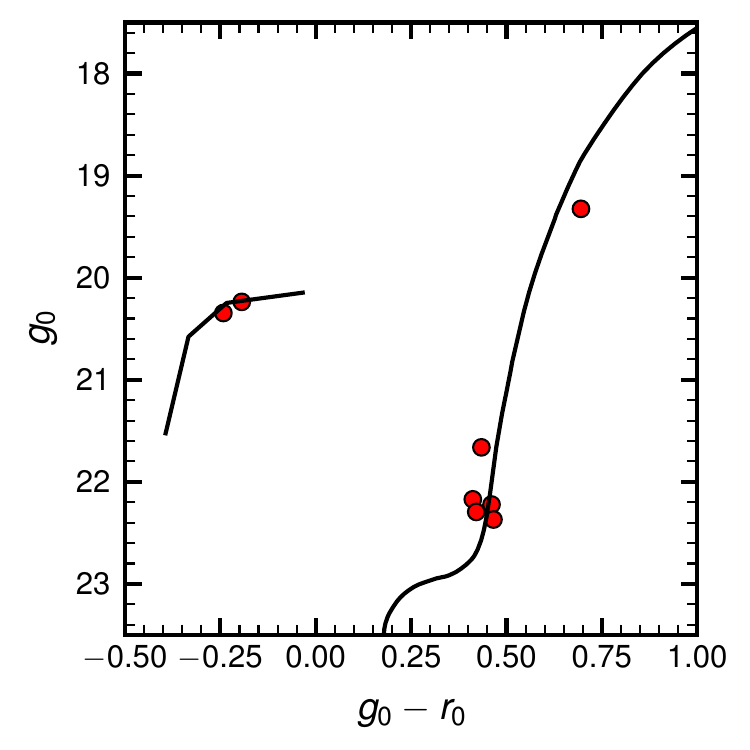}
\includegraphics[width=0.48\textwidth]{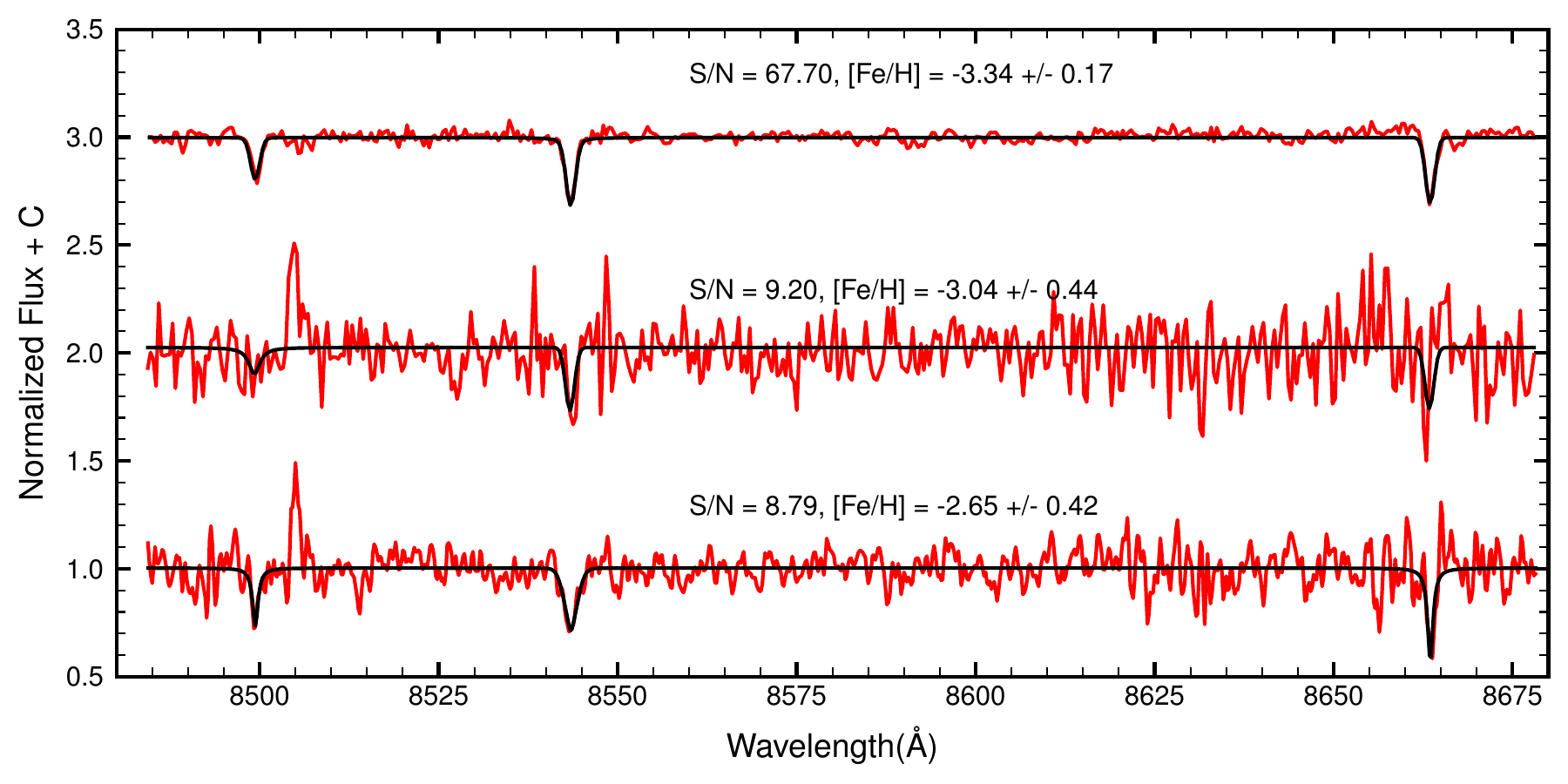}
\caption{(left) Spatial distribution of Eri~III member stars.  The dashed ellipses indicate 1, 3, and 5 times the half-light radius of the system, using structural parameters from \citet{conn18a}. (middle) Color-magnitude diagram of Eri~III, using DES DR2 photometry. Overplotted is an old (12.5~Gyr) metal-poor ($\feh=-2.5$, $[\alpha/\mathrm{Fe}]=0.4$) Dartmouth isochrone \citep{dotter08} at a heliocentric distance of 91~kpc \citep{conn18a}. As the Dartmouth isochrones do not include a horizontal branch, an empirical M92 ridge line is added for the BHB stars. (right) IMACS 600-line CaT spectra of Eri~III stars. The black curves show the best fit model with a Gaussian+Lorentzian fit to each of the CaT lines (see text for more details).  The top spectrum corresponds to the brightest RGB member at $g_0=19.33$, the middle spectrum corresponds to the second brightest RGB star at $g_0 = 21.66$, and the bottom spectrum is from a weighted average of the four faint RGB stars clumped near $g_0 \sim 22.3$. The S/N per pixel as well as the derived metallicity from the CaT EWs are shown for each spectrum.
\label{fig:eri3_imacs}}
\end{figure*}

We then selected all stars with measured velocities of $35~\kms < v_\mathrm{hel} < 70$~\kms\ from our two Eri~III slit masks as additional Eri~III candidates.  This selection yielded eight additional stars, two of which are far away from an old metal-poor isochrone at the distance of Eri~III and are therefore clear photometric non-members.  A third star, Gaia~DR3~4745738789619969920, has nearly identical colors and magnitudes as the brightest RGB member, but has much stronger metal lines in its spectrum, has a strongly discrepant proper motion from the initially identified members, and has a tentative parallax detection ($\omega/\sigma_{\omega}=2.8$) in the Gaia~DR3 catalog, suggesting that it is a foreground Milky Way star.  As illustrated in Fig.~\ref{fig:eri3_imacs}, the remaining five stars have magnitudes near $g=22$, are located along the Eri~III RGB, and appear to be metal-poor (see \S~\ref{sec:cat}), so we consider them likely members.  The only star for which we have multiple precise ($\sigma_{v} \lesssim3$~km~s$^{-1}$) velocity measurements is the brightest RGB star, and we do not detect any radial velocity variability over a time span of $\sim19$~months.  This star is therefore unlikely to be in a short-period binary system with a significant velocity amplitude, but we do not have the data to assess the impact of binarity on the remainder of the sample.

We used this eight star sample to constrain the internal kinematics of Eri~III.  With the maximum likelihood approach described by \citet{li17}, we found that the systemic velocity is $v_\mathrm{hel} = 53.7^{+1.7}_{-1.5}$~\kms.  Because the velocity uncertainties for all stars except the brightest RGB member are large ($\gtrsim5$~\kms), we are only able to place a weak upper limit on the velocity dispersion of Eri~III.  
Furthermore, given the small kinematic sample and large uncertainties, we found that the upper limit is sensitive to the choice of the prior as well as the boundaries of the prior. We therefore carefully selected the prior range and determined the upper limit on the velocity dispersion for both a uniform prior and log-uniform prior.  Assuming a conservative range of dynamical mass-to-light (M/L) ratios from 1--5000~$M_{\odot}/L_{\odot}$ (with the lower bound set by stellar population considerations and the upper bound set by the maximum value measured for dwarf galaxies; \citealt{wolf10}), the prior extends from 0.19--13.6~\kms.  With the uniform prior, we find a 90\%\ (95.5\%) upper limit of $\sigma < 9.1~(10.8)$~\kms, corresponding to $\textrm{M/L}_{V} < 2200~(3200)~M_{\odot}/L_{\odot}$ within the half-light radius, whereas with the log-uniform prior, the 90\%\ (95.5\%) upper limit is $\sigma < 5.4~(7.3)$~\kms, corresponding to $\textrm{M/L} < 790~(1440)~M_{\odot}/L_{\odot}$ within $r_{\mathrm{half}}$.\footnote{Extending the lower bound of the log-uniform prior down to 0.01~\kms results in a 90\%\ (95.5\%) upper limit of $\sigma < 3.3~(4.8)$~\kms, emphasizing the prior dependence of the result.}  In either case, substantial dark matter content in Eri~III cannot be excluded.

\subsubsection{DELVE 1}
\label{sec:delve1}

\begin{figure*}\centering
\includegraphics[width=0.24\textwidth]{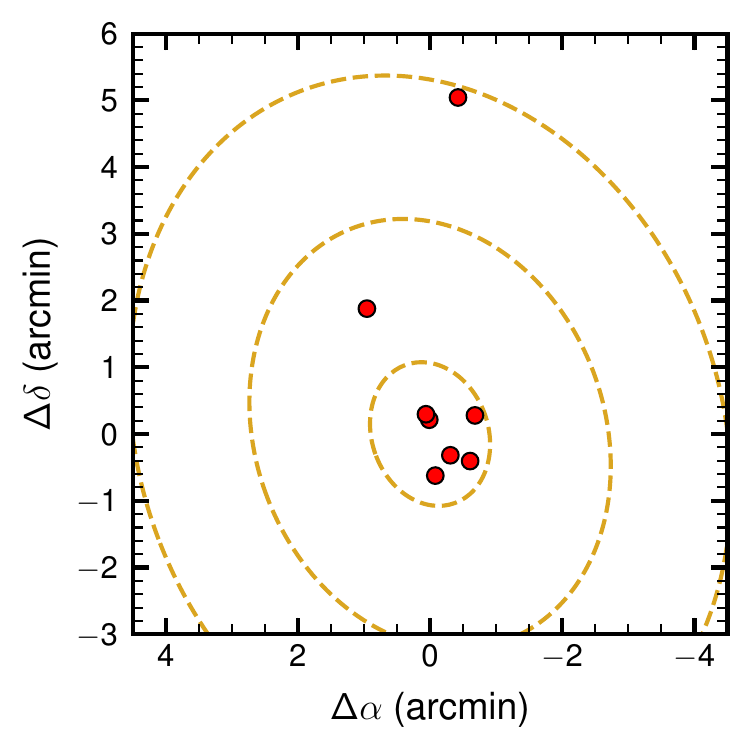}
\includegraphics[width=0.24\textwidth]{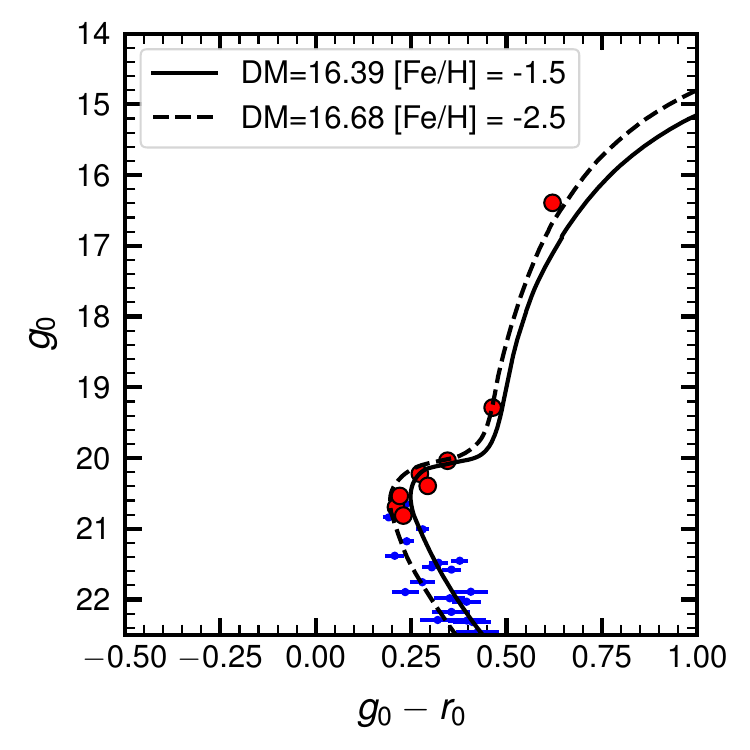}
\includegraphics[width=0.48\textwidth]{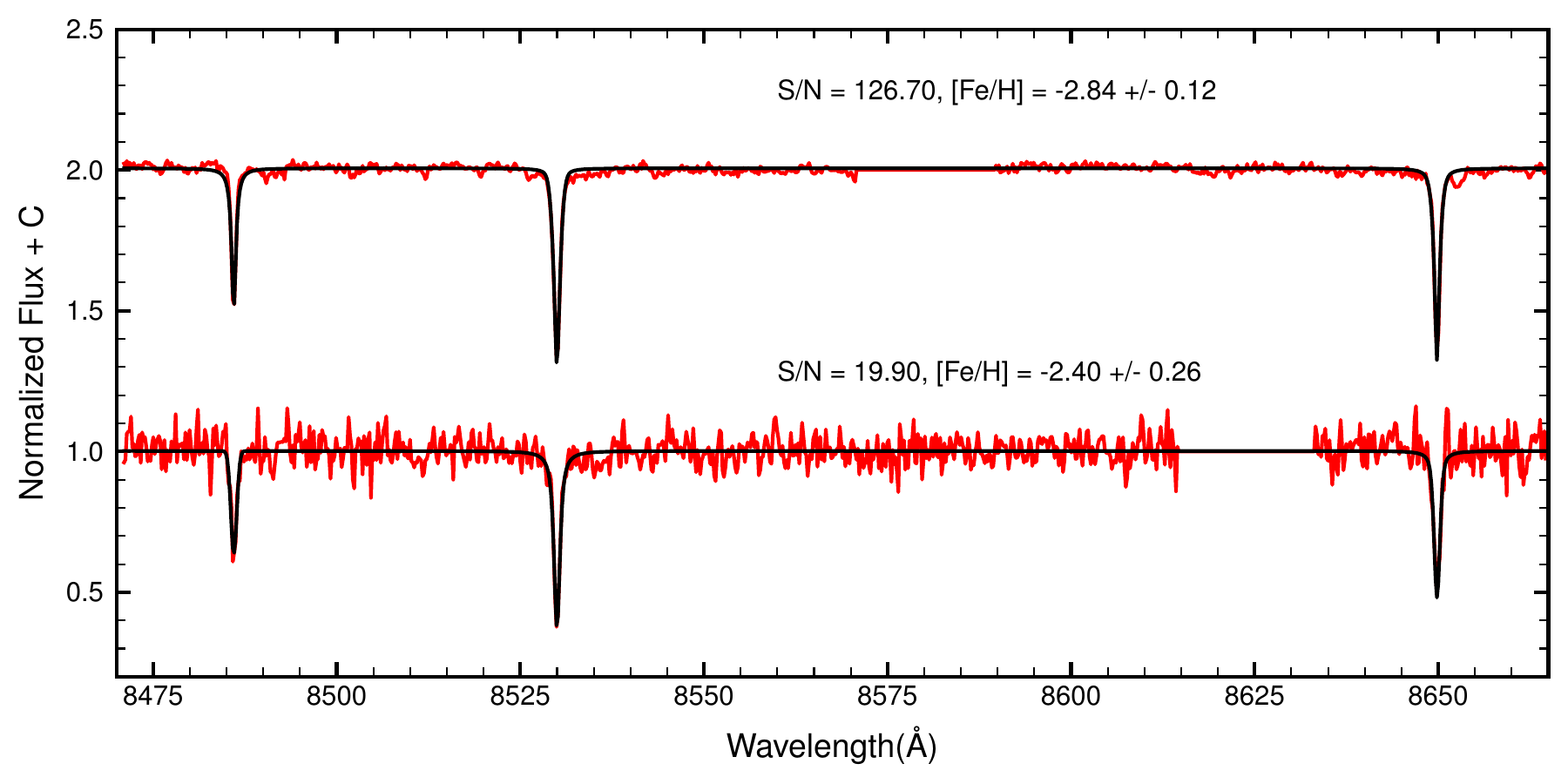}
\caption{(left) Spatial distribution of DELVE~1 member stars.  The dashed ellipses indicate 1, 3, and 5 times the half-light radius of the system, using structural parameters reported in the DELVE~1 discovery paper \citep{mau20}.   (middle) Color-magnitude diagram of DELVE~1, using DECaLS DR9 photometry. The spectroscopic members are displayed as red circles, and fainter candidate members within the half-light radius are plotted as blue dots with error bars. A Dartmouth isochrone \citep{dotter08} based on the parameters for DELVE~1 estimated in the discovery paper ($m-M = 16.39$~mag, $\feh = -1.5$, age~=~12.5~Gyr) is shown as the solid curve. However, using the spectroscopically confirmed member stars and a more appropriate lower-metallicity ($\feh = -2.5$) isochrone, we find that a larger distance modulus of $m-M = 16.68$~mag ($d = 22$~kpc, dashed curve) fits the data better, suggesting that DELVE~1 is farther away than was initially reported.
(right) IMACS spectra of the two brightest DELVE~1 members from the 2023 March observing run (top: $g_0=16.35$; bottom: $g_0=19.25$).  The black curves show the best fit model with a Gaussian+Lorentzian fit to each of the CaT lines.
\label{fig:delve1_imacs}}
\end{figure*}

The photometric member candidates for DELVE~1 cluster around a heliocentric velocity of $-400$~\kms, which is separated by more than 100~\kms\ from any foreground stars on the DELVE~1 slit mask.  The membership of DELVE~1 stars is therefore unambiguous.  Our IMACS data set includes eight DELVE~1 members, with two on the RGB and six on the subgiant branch or at the main sequence turnoff (see Fig.~\ref{fig:delve1_imacs}).  Making use of all spectra with $\textrm{S/N} > 3.8$, we have four observations of the brightest star, three observations of the next brightest four stars, and one observation each of the three faintest stars.  Six of the members are located within the half-light radius of DELVE~1, one is at $\sim2r_\mathrm{half}$, and one is at large radius ($\sim5r_\mathrm{half}$).  We find no evidence of velocity variability over 21 months for any of the stars with multiple observations, although only the two RGB members have small enough uncertainties for this test to be very meaningful.  We conclude that neither star is likely to be in a binary system with a period shorter than $\sim5$~years.

In the DELVE~1 discovery paper, the distance modulus was found to be $m-M = 16.39$~mag ($d=19$~kpc), with a relatively metal-rich isochrone at $\feh = -1.5$ providing the best fit \citep{mau20}. Because our IMACS spectroscopy shows that DELVE~1 is actually significantly more metal-poor, we re-fit the distance using a lower metallicity isochrone ($\feh = -2.5$).  For the new fit, we used a member sample combining the spectroscopically confirmed member stars with fainter photometric member candidates on the main sequence, as shown in the middle panel of Figure~\ref{fig:delve1_imacs}.  The photometric candidates were selected to be within the half-light radius of DELVE~1, within 0.3~mag in $g-r$ color of the metal-poor isochrone, and with $0.15 < g-r < 0.5$. We found that a distance modulus $m-M = 16.68$~mag (d = 22 kpc) maximizes the likelihood with the metal-poor isochrone, suggesting that DELVE 1 is farther away than was initially reported. 

After averaging together the repeat measurements of the brighter stars, we used the same maximum likelihood approach mentioned above \citep{li17} to characterize the stellar kinematics of DELVE~1.  We determined a systemic velocity of $v_\mathrm{hel} = -402.7 \pm 0.6$~\kms.  
As with Eri~III, the DELVE~1 velocity measurements do not resolve the velocity dispersion of the system.  Again based on the range of dynamical mass-to-light (M/L) ratios from 1--5000~$M_{\odot}/L_{\odot}$, we set a prior on the dispersion of 0.09--6.7~\kms.  With a uniform prior, we find a 90\%\ (95.5\%) upper limit of $\sigma < 2.5~(3.2)$~\kms, corresponding to $\textrm{M/L}_{V}<700~(1220)~M_{\odot}/L_{\odot}$ inside the half-light radius. With a log-uniform prior, we find a 90\%\ (95.5\%) upper limit of $\sigma < 1.2~(1.8)$~\kms ($\textrm{M/L}<160~[360]~M_{\odot}/L_{\odot}$ within $r_{\mathrm{half}}$).  Similar to Eri~III, these constraints do not allow us to rule out the possibility of dark matter in DELVE~1.  In both systems, the contribution of binary stars to the internal kinematics can only be constrained with additional velocity data.

With only two stars with metallicities that are 1.5$\sigma$ apart, we cannot place meaningful constraints on the metallicity distribution of DELVE~1.

\section{Chemical Abundance Analysis}
\label{sec:abundances}

\begin{figure*}
    \centering
    \includegraphics[width=\linewidth]{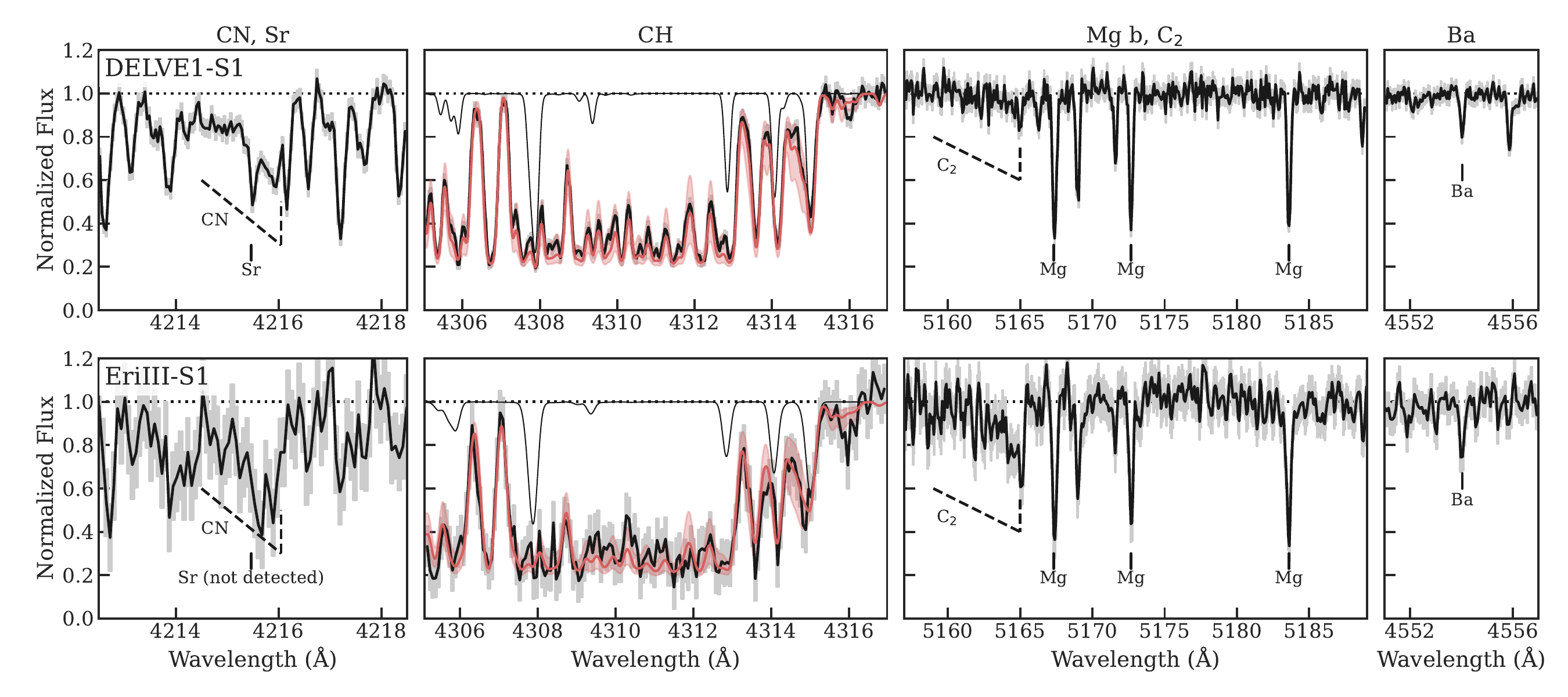}
    \caption{Cutouts of the MIKE spectra for DELVE~1-S1 and Eri~III-S1 around (from left to right) CN and Sr, CH, C$_{2}$ and Mg~b, and Ba features.
    The 1$\sigma$ spectrum uncertainties are shown as gray bars.
    The Sr line at 4215~{\AA} is not detected for Eri~III-S1, consistent with a low abundance inferred from the 4077~{\AA} line.
    On the CH panels, the thin black curve indicates a synthesis with no carbon, whereas the red curve and shaded red region indicate the best-fit synthesis and syntheses with the carbon abundance varied by $\pm 0.3$~dex.
    Both stars are highly enhanced in C and N but have low abundances of Sr and Ba.}
    \label{fig:MIKE_spectrum}
\end{figure*}

Figure~\ref{fig:MIKE_spectrum} shows portions of the MIKE spectra for the brightest DELVE~1 member star (Gaia~DR3~4359353258111283328, hereafter DELVE~1-S1) and the brightest Eri~III member star (Gaia~DR3~4745740262792352128, hereafter Eri~III-S1). It is immediately visually evident that these metal-poor stars are highly enhanced in carbon and nitrogen, with overall low neutron-capture element abundances. In this section, we describe the detailed chemical abundance analysis.

\subsection{Stellar Parameter Determination and Abundance Measurements}

Briefly, stellar parameters were determined from the MIKE spectra following \citet{Frebel2013}: first determining $\teff$, $\logg$, and $\nu_t$ by using \ion{Fe}{1} and \ion{Fe}{2} lines for excitation, ionization, and reduced equivalent width balance; then applying a temperature correction to bring $\teff$ to a photometric scale, and redetermining $\logg$ and $\nu_t$. This procedure produces stellar parameters consistent with those derived including non-local thermodynamic equilibrium effects \citep[e.g.,][]{Ezzeddine2020}. We verified using the \citet{Casagrande2010} color-temperature relations that the temperature is consistent within 100~K of photometrically derived values. The mean photometric temperature for different color-temperature relations is 85~K hotter for the DELVE~1 star and 105~K cooler for the Eri~III star, and the intrinsic scatter based on different photometric relations is 120~K. Propagating this uncertainty to other stellar parameters, we adopt overall systematic stellar parameter uncertainties of 120~K, 0.3~dex, 0.2~\kms, and 0.2~dex for effective temperature, surface gravity, microturbulence, and metallicity, respectively. The systematic stellar parameter uncertainties are added in quadrature to the statistical stellar parameter uncertainties \citep[e.g.,][]{ji20}. The parameters for both stars are listed in Table~\ref{tab:stellar_params}.

\begin{deluxetable*}{lcccc}
\tablecolumns{11}
\tabletypesize{\footnotesize}
\tablecaption{\label{tab:stellar_params}Stellar Parameters for MIKE Targets}
\tablehead{Name & $\teff$ & $\logg$ & $\nu_t$ & $\feh$ \\
{}&{(K)}&{}&{\kms}&{}}
\startdata
DELVE~1-S1 & $4950 \pm 135$ & $1.35 \pm 0.31$ & $2.06 \pm 0.23$ & $-2.83 \pm 0.22$ \\
Eri~III-S1 & $4880 \pm 222$ & $2.00 \pm 0.31$ & $2.08 \pm 0.28$ & $-3.10 \pm 0.37$
\enddata
\end{deluxetable*}

The analysis of the high-resolution spectra was performed with \texttt{smhr}\footnote{\url{https://github.com/andycasey/smhr}} \citep{Casey2014}, which wraps the ATLAS model atmospheres \citep{Castelli2003} and the MOOG 1D radiative transfer code with scattering and assuming local thermodynamic equilibrium \citep[LTE,][]{Sneden1973, sobeck11,Sobeck2023}\footnote{As implemented in \url{https://github.com/alexji/moog17scat}}. \texttt{smhr} provides an interface for radial velocity measurement, continuum normalization, equivalent width measurement, synthesis fitting, and error propagation. The overall analysis follows \citet{ji20}, except that we use an updated linelist from the $R$-Process Alliance (e.g., \citealt{roederer18}), using the atomic data from \texttt{linemake} \citep{Placco2021}\footnote{\url{https://github.com/vmplacco/linemake}}.
Abundances of most species (\ion{O}{1}, \ion{Na}{1}, \ion{Mg}{1}, \ion{K}{1}, \ion{Ca}{1}, \ion{Ti}{1}/\ion{Ti}{2}, \ion{Cr}{1}/\ion{Cr}{2}, \ion{Fe}{1}/\ion{Ti}{2}, and \ion{Ni}{1}) were determined with equivalent widths. Lines at $\lambda<4000$~\AA\ were all rejected for equivalent width measurements, as the low S/N in the blue inhibited accurate continuum placement. We synthesized the C-H band near 4300~\AA; the C-N bands near 3880~\AA\ and 4205~\AA; lines for \ion{Al}{1}, \ion{Si}{1}, and \ion{Sr}{2} due to blends; and lines for \ion{Sc}{2}, \ion{Co}{1}, and \ion{Ba}{2} due to hyperfine structure and/or isotopic splitting (using $r$-process ratios from \citealt{Sneden2008}). Formal $5\sigma$ upper limits for \ion{Eu}{2} were determined with spectral synthesis.
Abundance uncertainties for individual line measurements were determined by propagating the spectrum noise and stellar parameter uncertainties, with a minimum 0.1~dex systematic uncertainty on every line.
The final abundance is given by the inverse-variance weighted average of the line-by-line abundances.
The abundance results are listed in Table~\ref{tab:abunds}, scaled to solar abundances from \citet{Asplund2009}.
The individual line abundances are given in Table~\ref{tab:lines} (with atomic and molecular data sourced from \citealt{Lawler2001,Lawler2013,Lawler2015,Lawler2017,Lawler2019,DenHartog2011,DenHartog2014,DenHartog2019,DenHartog2021,DenHartog2023,Wood2013,Wood2014b,Wood2014a,Wood2018,Melendez2009,Ruffoni2014,Belmonte2017,PehlivanRhodin2017,Sobeck2007,Roederer2012,McWilliam1998,Sneden2014,Masseron2014,NIST}).
We note that all abundances are computed in LTE, which is sufficient for the qualitative comparative analysis in this paper to other metal-poor red giant stars. However, non-LTE effects can be larger than 0.2 dex for some elements (e.g., we expect the [Na/Fe] abundances to decrease by ${\sim}0.5$ dex due to the use of Na~D resonance lines; \citealt{Lind2011}) and should be applied before any detailed comparison.

\input{abundtab}

\input{linetab}

\subsection{Derived Abundances}

\begin{figure*}
    \centering
    \includegraphics[width=\linewidth]{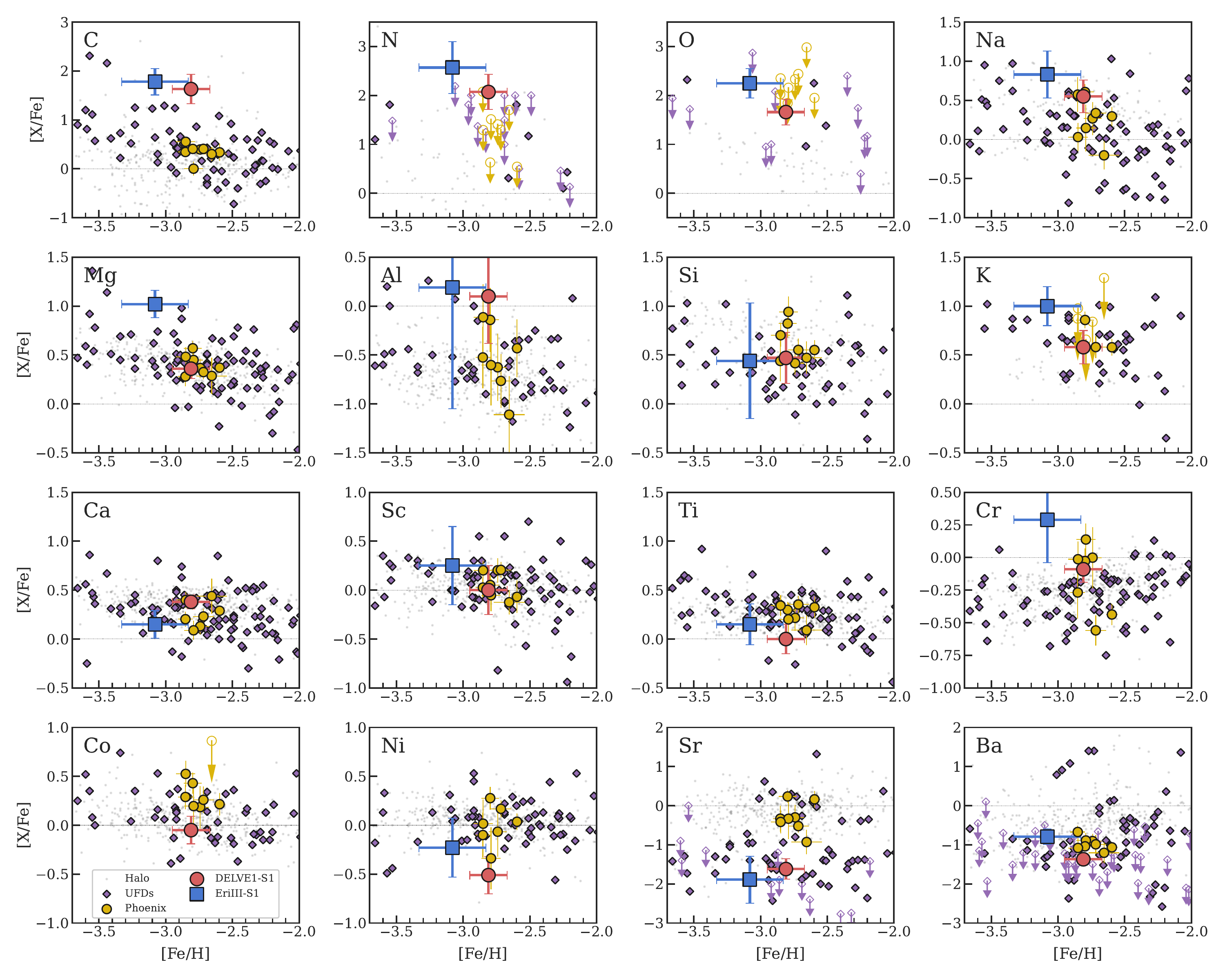}
    \caption{[X/Fe] vs. [Fe/H] for the brightest stars in DELVE~1 (red circle) and Eri~III (blue square) compared to Milky Way halo red giant stars (gray; $\log g < 3.5$ from \citealt{Barklem2005,Roederer2014b,Jacobson2015} as compiled in \citealt{jinabase}); ultra-faint dwarf galaxies \citep[purple diamonds;][open symbols with arrows indicate upper limits]{koch08,feltzing09,norris10,norris10b,frebel10,Simon2010,Roederer2014d,gilmore13,Koch2013,Frebel2014,ishigaki14,francois16,frebel16,ji16b,ji16c,roederer16,kirby17,hansen17,nagasawa18,chiti18,spite18,ji19,marshall19,ji20,hansen20,Chiti2023,Waller2023,Webber2023,hansen24}; and the Phoenix globular cluster stream \citep[][yellow circles]{Ji2020b,Casey2021}.
    The dotted horizontal line in each panel indicates $\mathrm{[X/Fe]} = 0$. For the Eri~III star, Co is not measured and Al and Si have sufficiently large error bars as to be considered unreliable.
    }
    \label{fig:xfegrid}
\end{figure*}

The abundance patterns of the two stars are presented in Fig.~\ref{fig:xfegrid}, where they are also compared with stars in the Milky Way halo and ultra-faint dwarfs.  The most notable features of both stars are their low metallicities and high carbon abundances.  We determine metallicities of $\feh = -3.08$ and $\feh = -2.81$ for Eri~III-S1 and DELVE~1-S1, respectively, consistent with the CaT metallicities.  Both stars have large overabundances of carbon, nitrogen, and oxygen ($\mbox{[C/Fe]} > 1.5$, $\mbox{[N/Fe]} \gtrsim 2$, $\mbox{[O/Fe]} > 1.5$). We note that O was measured from the weak 6300~\AA\ forbidden line and 7770~\AA\ triplet (when not blended with telluric lines). Though the latter have significant NLTE corrections, the fact that we detect these lines at all in metal-poor stars demonstrates that oxygen must be highly enhanced.

Eri~III also exhibits quite high abundances ($\mbox{[X/Fe]} \sim 1$) of the light and $\alpha$ elements Na, Mg, and K, whereas DELVE~1 abundances of these elements are in the expected range for metal-poor stars ($\mbox{[X/Fe]} \sim 0.5$).  The other $\alpha$ elements are normal (moderately enhanced relative to solar) for both stars. We report Al abundances, though the uncertainties are sufficiently large that they should be largely disregarded.

The iron peak elements are mostly normal in these stars. An apparently high abundance of Cr in Eri~III can be attributed mostly to spectrum noise, as it was derived from just one line.
Mn is normally measurable in similar stars, but only the 4030~\AA\ triplet was detectable, which we rejected due to large NLTE effects \citep{Bergemann2019}.
However, it is notable that the Ni abundances are low in both stars, especially in the DELVE~1 star with $\mbox{[Ni/Fe]} = -0.5$.

Both stars contain very low levels of neutron-capture elements,  with $\mbox{[Sr/Fe]} \sim -1.6$ and $\mbox{[Ba/Fe]} = -1.4$ for DELVE~1-S1 and $\mbox{[Sr/Fe]} \sim -1.9$ and $\mbox{[Ba/Fe]} = -0.8$ for Eri~III-S1.
We note that a precise Sr abundance measurement in Eri~III-S1 is hindered by the very low S/N ratio at the blue end of the spectrum, but the abundance is clearly very low.

\subsection{Stellar Classification}

Based on their high carbon abundances ($\mbox{[C/Fe]} > 1$) and low neutron-capture content ($\mbox{[Ba/Fe]} < 0$), both stars observed at high spectral resolution can be classified as CEMP-no stars \citep[e.g.,][]{bc05,carollo14}.  The two stars are also close to the defining criteria for the rare class of nitrogen-enhanced metal-poor (NEMP) stars ($\mbox{[N/Fe]} > 0.5$ and $\mbox{[C/N]} < -0.5$; \citealt{johnson07}).  The high nitrogen abundances are presumed to result from hot bottom burning in high mass asymptotic giant branch (AGB) stars, which may transfer enriched material to a binary companion to produce a low mass, metal-poor, and nitrogen-rich NEMP star \citep[e.g.,][]{pols12}.  Some NEMP stars share the high Na and Mg abundances of Eri~III-S1 and DELVE~1-S1, although they also exhibit much higher Ba abundances, as expected from $s$-process nucleosynthesis in AGB stars \citep{pols12}.  It is therefore not clear whether these stars should be considered NEMP stars.

\section{Interpretation}
\label{sec:interp}

\subsection{Comparison with Globular Clusters}

Carbon-enriched stars are quite rare in globular clusters, and are generally either found on the AGB or thought to have been polluted with the carbon-rich nucleosynthetic products from a companion star that passed through the AGB phase \cite[e.g.,][]{harding62,bond75,cote97,sharina12,kirby15c}, resulting in a CEMP-s classification.  As far as we are aware, there are no confirmed CEMP-no stars associated with globular clusters.

\begin{figure}
    \centering
    \includegraphics[width=1.0\linewidth]{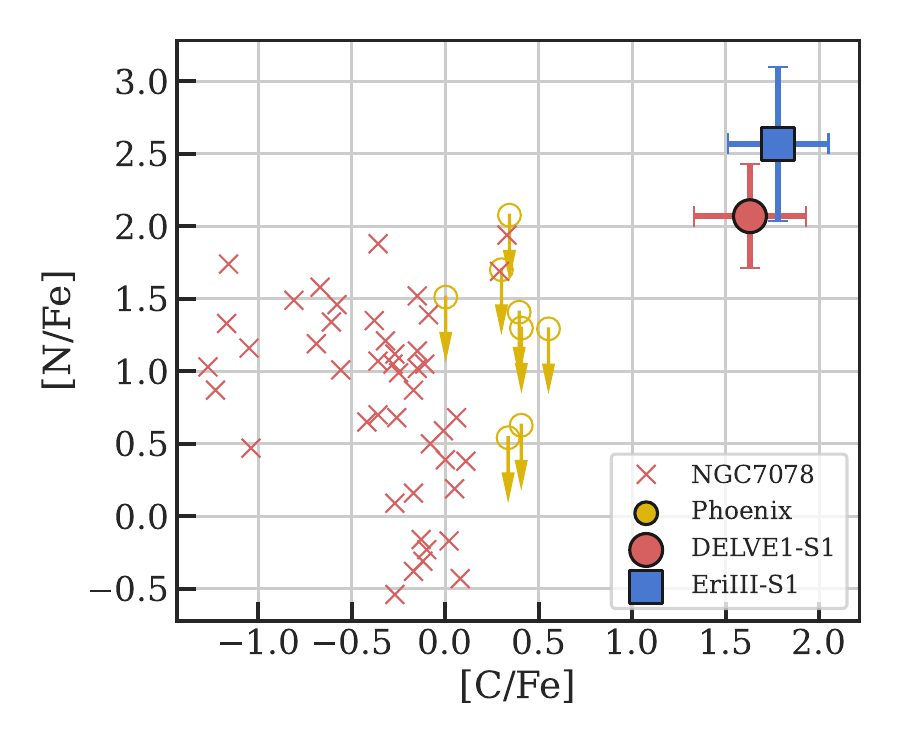}
    \caption{C and N abundances of DELVE~1 and Eri~III compared to NGC~7078 \citep{Roediger2014} and the Phoenix stream \citep{Ji2020b,Casey2021}. Both C and N are high in our target stars, and thus clearly inconsistent with the C-N anticorrelation typically seen in multiple populations within globular clusters (even though Phoenix only has upper limits for [N/Fe], these limits are sufficient to establish that its N abundances are substantially lower than the Eri~III and DELVE~1 stars). The same is true for the Na-O anticorrelation, except that there are only upper limits for O (not shown).}
    \label{fig:CN}
\end{figure}

The Eri~III and DELVE~1 stars also exhibit high nitrogen abundances.  In massive globular clusters, carbon and nitrogen abundances are typically anti-correlated, as illustrated for the examples of NGC~7078 and the Phoenix Stream (which originated from a globular cluster; \citealt{Ji2020b}) in Fig.~\ref{fig:CN}.  Only a single NEMP star has been identified in a globular cluster  \citep[ESO280-SC06;][]{sm19}, but the abundances of elements other than carbon and nitrogen have not been measured in that star, so we cannot assess how much it resembles Eri~III-S1 and DELVE~1-S1.

The low metallicities of the Eri~III and DELVE~1 stars stand out from the globular cluster population as well.  The Milky Way lacks any globulars at $\feh \lesssim -2.5$ \citep[e.g.,][]{harris10,sobeck11,lovisi13,kirby23}, although two streams from disrupted clusters that are more metal-poor have recently been discovered \citep{wan20,Casey2021,martin22,yuan22}, and \citet{larsen20} identified a cluster at $\feh = -2.9$ in M31.  Thus, even though it is evidently possible for clusters to form at the metallicity of Eri~III ($\feh = -3.08$) and DELVE~1 ($\feh = -2.81$), the Milky Way does not appear to contain any such objects that have survived to the present day.

\subsection{Comparison with Dwarf Galaxies}

\begin{figure*}
    \centering
    \includegraphics[width=1.0\textwidth]{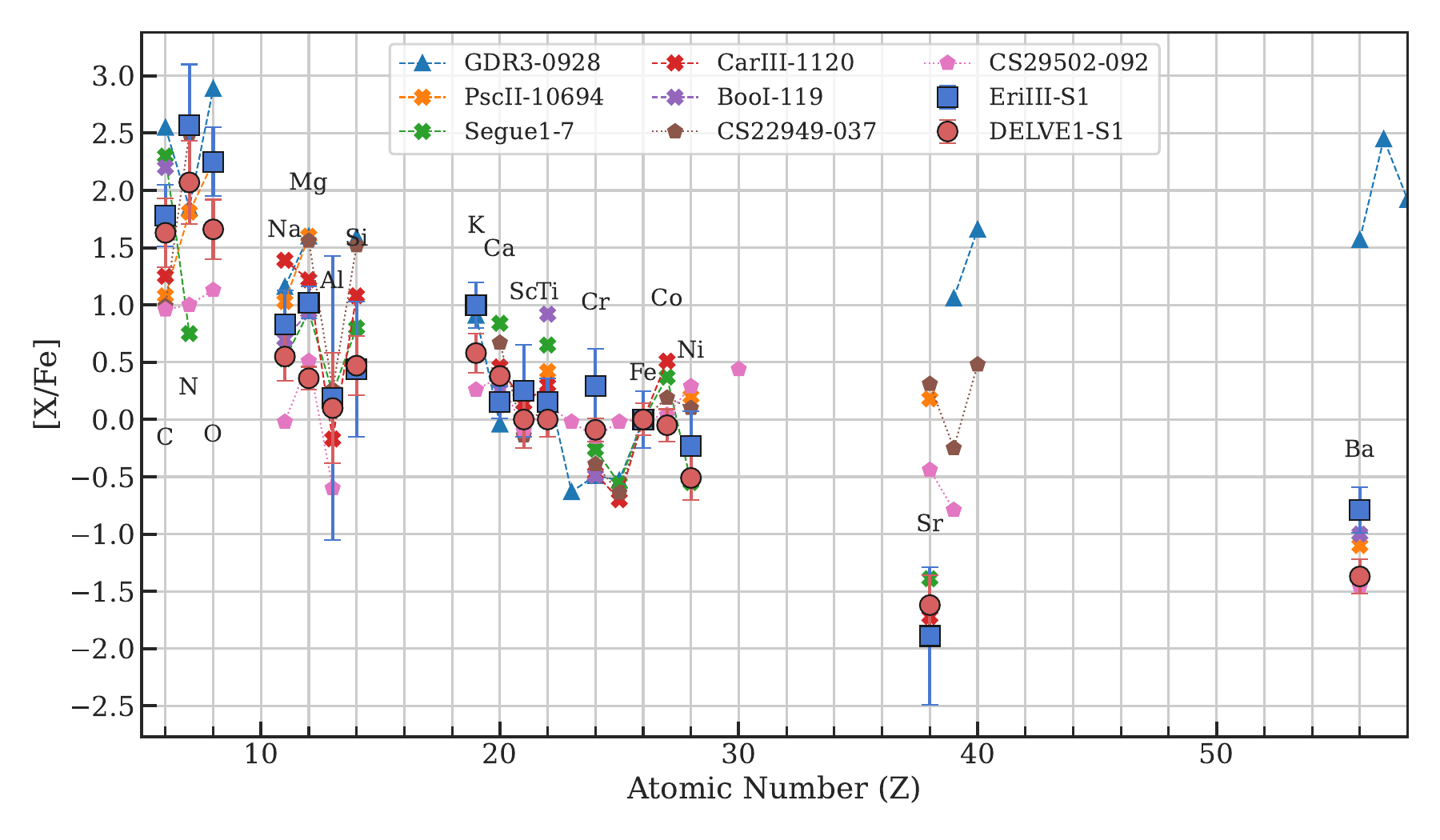}
    \caption{[X/Fe] vs atomic number for the DELVE~1 and Eri~III stars.
    They are compared to abundances of other CEMP stars with $\mbox{[C/Fe]} > 1$ in ultra-faint dwarfs:
    GDR3-0928 \citep[Ret~II,][]{hayes23}, PscII-10694 \citep{spite18}, Segue1-7 \citep{norris10c}, CarIII-1120 \citep{ji20}, BooI-119 \citep{gilmore13,lai11}, and two CEMP halo stars \citep[CS22949-037 and CS29502-092,][]{Roederer2014b}.
    }
    \label{fig:xfeZ}
\end{figure*}

Conversely, neither the metallicities nor the carbon abundances of the Eri~III and DELVE~1 stars would be unusual among ultra-faint dwarfs.  CEMP-no stars have also been identified in Ursa~Major~II, Segue~1, Tucana~II, Bo{\"o}tes~I, Pisces~II, Carina~III, and Tucana~V \citep[][]{frebel10,norris10c,gilmore13,chiti18,spite18,ji20,hansen24}, as well as in several of the classical dwarf spheroidals \citep{skuladottir20,skuladottir24,hansen23,roederer23}.  Many of the ultra-faint dwarf CEMP-no stars also share the high [Na/Fe] and [Mg/Fe] abundances observed in Eri~III, again indicating consistency with the dwarf galaxy population.  In Fig.~\ref{fig:xfeZ}, we compare the [X/Fe] measurements for the Eri~III and DELVE~1 stars with those of dwarf galaxy CEMP (mostly CEMP-no) stars, finding generally good agreement.

The chemical abundance patterns of their brightest stars are therefore suggestive of a dwarf galaxy classification for Eri~III and DELVE~1.  These systems have half-light radii that are at least a factor of 3 smaller than the most compact previously confirmed Milky Way dwarf \citep{munoz18}, which would significantly expand the parameter space occupied by faint galaxies.  One possible mechanism that could account for the small sizes of these systems is tidal stripping.  \citet{errani23} showed that if the dark matter halo and the initial stellar distribution of a dwarf galaxy are both cuspy, the system will not be completely destroyed by tidal forces, and after sufficient stripping it can be reduced to a half-light radius and luminosity comparable to Eri~III and DELVE~1.  However, reaching this micro galaxy status requires stripping by a factor of $\sim10^{4}$.  For a dwarf galaxy that initially formed on the luminosity-metallicity relation from \citet{kirby13b}, reducing its luminosity by such a large amount should leave the surviving remnant $>1$~dex more metal-rich than the relation.  Since Eri~III and DELVE~1 have metallicities roughly consistent with the luminosity-metallicity relation for their current luminosities, if they have lost 99.99\% of their stars via tidal stripping then their progenitors must have started off $>1$~dex more metal-poor than the relation.  It is thus difficult to reconcile the observations of these objects with the tidal stripping hypothesis if they are dwarf galaxies.  Regardless of the physical explanation for their sizes, if satellites as small as $\sim5$~pc are indeed galaxies \citep[e.g.,][]{ricotti16}, the number of Milky Way dwarfs may be substantially larger than currently recognized \citep[e.g.,][]{kim18,nadler20b,mk22}.

\subsection{The Nature of Eri~III and DELVE~1}

Despite the closer chemical resemblance with dwarf galaxies, none of the available data on either object actually requires the operation of physics beyond that of baryons and Newton's laws of gravity.  Under the \citet{ws12} definition of a galaxy, then, Eri~III and DELVE~1 do not clearly qualify according to our present knowledge.  If one or both objects are in fact dark matter-free star clusters, their chemical properties are completely different from those of classical globular clusters, implying a distinct formation mechanism.  One hypothesis for the origin of CEMP-no stars at extremely low metallicities is that they are the product of faint supernova explosions in which most of the heavy elements fall back onto the compact object and only the light elements are ejected in large quantities \citep[e.g.,][]{un03,ishigaki14b,chansen16,frebel19}.  We suggest that Eri~III and DELVE~1 could be examples of small clusters that formed immediately after such a Population III supernova and were enriched primarily or exclusively by that event.  Because of their low stellar masses, they are unlikely to have hosted a second generation of star formation that could be enriched by the AGB ejecta of the intermediate mass stars from the first generation, as is thought to have occurred in larger clusters \citep[e.g.,][]{carretta10,bl18}.  In this scenario, objects like Eri~III and DELVE~1 would be the most primitive clusters ever formed, and perhaps the only stellar systems in which the nucleosynthetic products of the first supernovae are so cleanly preserved.

If Eri~III and DELVE~1 are clusters, it is possible that they formed within low-luminosity dwarf galaxies.  Since neither object is surrounded by a visible galaxy today, the hosts must have been tidally disrupted, leaving behind just their clusters and perhaps a faint (not yet observed) stream.  The prototypes for clusters within ultra-faint dwarfs are the central star cluster in Eridanus~II \citep{koposov15,crnojevic16,simon21,alzate21,martinezvazquez21,weisz23} and the cluster in Ursa~Major~II \citep{zucker06,eadie22}.  The UMa~II cluster is extremely poorly studied, with neither any published spectroscopy nor an analysis of its stellar population, so nothing can currently be said about its chemical properties.  Its physical size is similar to Eri~III and DELVE~1 \citep{eadie22}, whereas the Eri~II cluster is a factor of $\sim2$ larger \citep{simon21}.  The cluster in Eri~II has been the subject of several photometric studies, which concluded that it is old and metal-poor ($\feh \sim -2.5$), with properties very similar to those of Eri~II itself \citep[e.g.,][]{simon21,weisz23}.  With its brightest stars at $V \approx 22$, spectroscopy of this object is challenging.  Nevertheless, \citet{zoutendijk20} obtained low-resolution MUSE spectra covering seven candidate cluster members, finding that most of them have metallicities of $\feh \approx -1.5$, with one star possibly much more metal-poor.  The Eri~II cluster thus may have a substantially higher metallicity than Eri~III and DELVE~1.  If the metallicity of a cluster is correlated with the metallicity of its host galaxy, this measurement would suggest that clusters with the metallicities of Eri~III and DELVE~1 originated in lower-mass dwarfs than Eri~II.  In the cluster hypothesis, the CEMP-no chemical signatures of Eri~III and DELVE~1 are likely related to their unique formation environment.  Evidence in favor of this idea could be obtained via additional spectroscopy of the UMa~II and Eri~II clusters to search for a similar chemical signature, or by detection of stellar streams along the orbits of Eri~III and DELVE~1.

\subsection{Observationally Distinguishing the Smallest Dwarfs from Clusters}

The classification of Eri~III and DELVE~1 as either the smallest dwarf galaxies or the first primordial star clusters discovered has important implications for the satellite population of the Milky Way and early star and cluster formation.  These two possibilities leave us facing a key question: how can we determine whether these objects are galaxies or clusters?

If Eri~III and DELVE~1 are purely stellar systems, then according to the \citet{wolf10} formula their velocity dispersions would be $\sim0.3$~km~s$^{-1}$ and $\sim0.1$~km~s$^{-1}$, respectively (for a stellar mass-to-light ratio of $2~M_{\odot}/L_{\odot}$).  These values are likely to remain unmeasurably small for the foreseeable future.  Alternatively, masses within the half-light radius of $\gtrsim10^{5}~M_{\odot}$, as is typical for most ultra-faint dwarfs, would lead to expected velocity dispersions of $\gtrsim4$~km~s$^{-1}$, which could be measurable via deep spectroscopy with Keck/DEIMOS \citep[e.g.,][]{simon11} or VLT/FLAMES \citep[e.g.,][]{jenkins21}.  Our upper limit on the velocity dispersion of Eri~III is consistent with a mass this large.  The constraint we have already obtained on the velocity dispersion of DELVE~1 rules out such a high mass.  However, since Eri~III and DELVE~1 have smaller half-light radii than confirmed UFDs, their half-light masses would be smaller even if they inhabit similar dark matter halos.  Masses on the order of $M_{\rm half} \sim 10^{4}~M_{\odot}$, corresponding to $\sigma \sim 1.5$~km~s$^{-1}$ and $M/L \sim 100~M_{\odot}/L_{\odot}$, cannot currently be excluded. 

The recently discovered satellite Ursa~Major~III/UNIONS~1 \citep{smith24} offers an instructive example.  Ursa~Major~III/UNIONS~1 is several magnitudes fainter than Eri~III or DELVE~1, with a half-light radius of just $3 \pm 1$~pc.  Nevertheless, the system has a tentatively measured velocity dispersion of $\sim2-4$~\kms, depending on which stars are included in the calculation \citep{smith24}, which would imply a large dark matter content \citep{errani24}.  Binary stars may be influencing this measurement, so additional observations are needed to confirm whether the dispersion actually reflects the dynamical mass of Ursa~Major~III/UNIONS~1.  If this even fainter and more compact system can be confirmed as a dwarf, that would provide a strong motivation for further pursuing the kinematics of Eri~III and DELVE~1.

Perhaps a more promising approach for conclusively classifying Eri~III and DELVE~1 relies on chemical abundance measurements for additional stars.  If they are actually clusters, then we expect that they should be chemically homogeneous, and given the abundances of Eri~III-S1 and DELVE~1-S1, all of the stars should be carbon-enhanced.  $\mbox{[C/Fe]} \sim 1.5$ should be an easy signature to identify observationally even for faint stars.  Perhaps supporting this idea for DELVE~1, we note that the photometric member sample shown in the middle panel of Fig.~\ref{fig:delve1_imacs} tends to lie slightly redward of the best-fit isochrone, as might be expected if the stars are carbon-rich.  However, given the small number of stars, the uncertainties in the stellar models, and the lack of spectroscopic confirmation, this idea is only speculative at the moment.  On the other hand, dwarf galaxy stars should exhibit a range in metallicity, as well as likely lower carbon abundances for most of the stars.  The former could be measured either via low-resolution Ca~K spectroscopy \citep{chiti18b} or narrow-band photometry \citep[e.g.,][]{fu22,fu23}, whereas the latter would require medium-resolution spectroscopy.

Finally, stellar mass segregation as a result of dynamical relaxation may offer an additional means of distinguishing faint dwarf galaxies and clusters \citep[e.g.,][]{kim15kim2}.  \citet{baumgardt22} used HST imaging of Eri~III to compare the radial profiles of $\sim0.53$~M$_{\odot}$ and $\sim0.78$~M$_{\odot}$ stars, finding that the ratio of the extent of the more massive stars to the less massive ones is $0.84 \pm 0.09$.  From this measurement, they estimate a probability of mass segregation of 91.7\%, which is suggestive of a cluster but not definitive.  Deeper observations could improve this result.  No mass segregation analysis has been performed on DELVE~1 to date.

\section{Summary}\label{sec:summary}

One of the present challenges in the observational study of low-luminosity Milky Way satellites is identifying which of the objects discovered over approximately the past decade are dwarf galaxies and which are not.  We have presented the first spectroscopy of two ultra-faint stellar systems, Eri~III and DELVE~1.  We measured the velocities and metallicities of the brightest several stars in each satellite, showing that both Eri~III and DELVE~1 are metal-poor and placing upper limits on their stellar velocity dispersions.

We also obtained high-resolution spectroscopy of the single brightest star in each system and performed a chemical abundance analysis.  In both cases, the star has a metallicity near the extremely metal-poor boundary ($\feh = -2.81$ for DELVE~1-S1 and $\feh = -3.08$ for Eri~III-S1) and is strongly carbon-enhanced.  The two stars exhibit low abundances of neutron-capture elements, meriting a classification as CEMP-no stars.

The observed chemical abundance patterns are puzzling in light of the very small sizes of Eri~III and DELVE~1, and their resulting presumptive classification as clusters.  We explore two possible hypotheses for the nature of these systems: (1) they are dwarf galaxies, in which case the chemical abundances are not unusual, but they are at least a factor of 3 more compact than any previously known faint dwarfs, or (2) they are carbon-enhanced extremely metal-poor star clusters, chemically distinct from all other known clusters.  Supporting the dwarf galaxy picture, \citet{fu23} photometrically measured a metallicity spread within Eri~III at almost $3\sigma$ confidence.  On the other hand, \citet{baumgardt22} detected mass segregation among the Eri~III stars at close to $2\sigma$ significance, suggesting the absence of dark matter.  Further spectroscopy of Eri~III and DELVE~1 is needed in order to resolve this issue.

Confirming either scenario for the classification of Eri~III and DELVE~1 would be an exciting result.  If these objects are dwarf galaxies, then the same may be true for many of the other recently discovered compact ultra-faint satellites, significantly increasing the retinue of dwarfs surrounding the Milky Way.  If instead they are clusters, then they would represent a qualitatively new regime of cluster chemical evolution, potentially providing new insight into early nucleosynthesis.  We encourage spectroscopy of these two systems, as well as other unclassified compact ultra-faint stellar systems, in order to determine which possibility is correct.  More detailed studies of the two known examples of clusters that formed within ultra-faint dwarf galaxies, in Eri~II and UMa~II, could also shed light on how such clusters might differ from those originating in more massive galaxies.

\section{Acknowledgements}\label{sec:ack}

We thank the anonymous referee for comments that improved the paper.

T.S.L. acknowledges financial support from Natural Sciences and Engineering Research Council of Canada (NSERC) through grant RGPIN-2022-04794.  A.P.J. acknowledges support by the National Science Foundation under grants AST-2206264 and AST-2307599. 
A.B.P. acknowledges supported from NSF grant AST-1813881.
T.T.H acknowledges support from the Swedish Research Council (VR 2021-05556).
W.C gratefully acknowledges support from a Gruber Science Fellowship at Yale University.
S.E.K. acknowledges support from Science \& Technology
Facilities Council (STFC) (grant ST/Y001001/1).  
A.D.W. acknowledges support by the National Science Foundation under grants AST-2006340, AST-2108168, and AST-2307126. 
E.N.K.\ acknowledges support from NSF CAREER grant AST-2233781.

This research has made use of NASA's Astrophysics Data System
Bibliographic Services and the local volume database\footnote{\url{https://github.com/apace7/local_volume_database}}.

For the purpose of open access, the authors have applied a Creative Commons Attribution (CC BY) licence to any Author Accepted Manuscript version arising from this submission. 

This project used public archival data from the Dark Energy Survey (DES). Funding for the DES Projects has been provided by the U.S. Department of Energy, the U.S. National Science Foundation, the Ministry of Science and Education of Spain, the Science and Technology Facilities Council of the United Kingdom, the Higher Education Funding Council for England, the National Center for Supercomputing Applications at the University of Illinois at Urbana-Champaign, the Kavli Institute of Cosmological Physics at the University of Chicago, the Center for Cosmology and Astro-Particle Physics at the Ohio State University, the Mitchell Institute for Fundamental Physics and Astronomy at Texas A\&M University, Financiadora de Estudos e Projetos, Funda{\c c}{\~a}o Carlos Chagas Filho de Amparo {\`a} Pesquisa do Estado do Rio de Janeiro, Conselho Nacional de Desenvolvimento Cient{\'i}fico e Tecnol{\'o}gico and the Minist{\'e}rio da Ci{\^e}ncia, Tecnologia e Inova{\c c}{\~a}o, the Deutsche Forschungsgemeinschaft, and the Collaborating Institutions in the Dark Energy Survey.

The Collaborating Institutions are Argonne National Laboratory, the University of California at Santa Cruz, the University of Cambridge, Centro de Investigaciones Energ{\'e}ticas, Medioambientales y Tecnol{\'o}gicas-Madrid, the University of Chicago, University College London, the DES-Brazil Consortium, the University of Edinburgh, the Eidgen{\"o}ssische Technische Hochschule (ETH) Z{\"u}rich,  Fermi National Accelerator Laboratory, the University of Illinois at Urbana-Champaign, the Institut de Ci{\`e}ncies de l'Espai (IEEC/CSIC), the Institut de F{\'i}sica d'Altes Energies, Lawrence Berkeley National Laboratory, the Ludwig-Maximilians Universit{\"a}t M{\"u}nchen and the associated Excellence Cluster Universe, the University of Michigan, the National Optical Astronomy Observatory, the University of Nottingham, The Ohio State University, the OzDES Membership Consortium, the University of Pennsylvania, the University of Portsmouth, SLAC National Accelerator Laboratory, Stanford University, the University of Sussex, and Texas A\&M University.

Based in part on observations at Cerro Tololo Inter-American Observatory, National Optical Astronomy Observatory, which is operated by the Association of Universities for Research in Astronomy (AURA) under a cooperative agreement with the National Science Foundation.

This work has made use of data from the European Space Agency (ESA) mission
{\it Gaia} (\url{https://www.cosmos.esa.int/gaia}), processed by the {\it Gaia}
Data Processing and Analysis Consortium (DPAC,
\url{https://www.cosmos.esa.int/web/gaia/dpac/consortium}). Funding for the DPAC
has been provided by national institutions, in particular the institutions
participating in the {\it Gaia} Multilateral Agreement.

The Legacy Surveys consist of three individual and complementary projects: the Dark Energy Camera Legacy Survey (DECaLS; Proposal ID \#2014B-0404; PIs: David Schlegel and Arjun Dey), the Beijing-Arizona Sky Survey (BASS; NOAO Prop. ID \#2015A-0801; PIs: Zhou Xu and Xiaohui Fan), and the Mayall z-band Legacy Survey (MzLS; Prop. ID \#2016A-0453; PI: Arjun Dey). DECaLS, BASS and MzLS together include data obtained, respectively, at the Blanco telescope, Cerro Tololo Inter-American Observatory, NSF’s NOIRLab; the Bok telescope, Steward Observatory, University of Arizona; and the Mayall telescope, Kitt Peak National Observatory, NOIRLab. Pipeline processing and analyses of the data were supported by NOIRLab and the Lawrence Berkeley National Laboratory (LBNL). The Legacy Surveys project is honored to be permitted to conduct astronomical research on Iolkam Du’ag (Kitt Peak), a mountain with particular significance to the Tohono O’odham Nation.

NOIRLab is operated by the Association of Universities for Research in Astronomy (AURA) under a cooperative agreement with the National Science Foundation. LBNL is managed by the Regents of the University of California under contract to the U.S. Department of Energy.

This project used data obtained with the Dark Energy Camera (DECam), which was constructed by the Dark Energy Survey (DES) collaboration. 

BASS is a key project of the Telescope Access Program (TAP), which has been funded by the National Astronomical Observatories of China, the Chinese Academy of Sciences (the Strategic Priority Research Program “The Emergence of Cosmological Structures” Grant \# XDB09000000), and the Special Fund for Astronomy from the Ministry of Finance. The BASS is also supported by the External Cooperation Program of Chinese Academy of Sciences (Grant \# 114A11KYSB20160057), and Chinese National Natural Science Foundation (Grant \# 12120101003, \# 11433005).

The Legacy Survey team makes use of data products from the Near-Earth Object Wide-field Infrared Survey Explorer (NEOWISE), which is a project of the Jet Propulsion Laboratory/California Institute of Technology. NEOWISE is funded by the National Aeronautics and Space Administration.

The Legacy Surveys imaging of the DESI footprint is supported by the Director, Office of Science, Office of High Energy Physics of the U.S. Department of Energy under Contract No. DE-AC02-05CH1123, by the National Energy Research Scientific Computing Center, a DOE Office of Science User Facility under the same contract; and by the U.S. National Science Foundation, Division of Astronomical Sciences under Contract No. AST-0950945 to NOAO.

\facility{Magellan:I (IMACS); Magellan:II (MIKE)}

\bibliography{main_jds}{}
\bibliographystyle{aasjournal}

\end{document}

%% file: imacs_spec_table_membersonly.tex

\tabletypesize{\scriptsize}
\begin{deluxetable*}{c c c c c c c r c r c c c}
\tablecaption{IMACS Velocity and Metallicity Measurements for Eri~III and DELVE~1.\label{tab:imacs_spec_table}}

\tablehead{ID\tablenotemark{a} & RA & Dec & $g$\tablenotemark{b} & $r$\tablenotemark{b} & $z$\tablenotemark{b} & MJD & Grating & S/N & \multicolumn{1}{c}{$v$\tablenotemark{c}} & ${\rm EW}$ & ${\rm [Fe/H]}$ & Mem\tablenotemark{d} \\ 
 & (deg) & (deg) & (mag) & (mag) & (mag) &  & &  & \multicolumn{1}{c}{(\kms)} & (\AA) &  &  }
\startdata
DES\,J022243.71$-$521715.0  &   35.68215  & $ -52.28751 $ &  21.74  &  21.28  &  21.02  &  59471.0  &   G600  &    9.2  &   \phs$47.3 \pm 5.8$  &  $1.22 \pm 0.47$  &  $-3.02 \pm 0.44$  &  1 \\
DES\,J022243.82$-$521655.9  &   35.68260  & $ -52.28219 $ &  22.38  &  21.93  &  21.74  &  59471.0  &   G600  &    5.4  &   \phs$60.4 \pm 8.7$  &         ...       &         ...        &  1 \\
       4745740335808220800  &   35.68838  & $ -52.27641 $ &  20.32  &  20.49  &  20.66  &  58882.1  &  G1200  &    3.5  &   \phs$61.8 \pm 7.0$  &         ...       &         ...        &  1 \\
                            &   35.68838  & $ -52.27641 $ &  20.32  &  20.49  &  20.66  &  59471.0  &   G600  &   14.5  &   \phs$61.5 \pm 4.8$  &         ...       &         ...        &  1 \\
DES\,J022245.46$-$521706.8  &   35.68940  & $ -52.28522 $ &  22.45  &  21.96  &  21.67  &  59471.0  &   G600  &    5.2  &  \phs$40.6 \pm 10.7$  &         ...       &         ...        &  1 \\
       4745740262792353536  &   35.69377  & $ -52.28267 $ &  20.43  &  20.64  &  20.89  &  58882.1  &  G1200  &    2.9  &   \phs$63.2 \pm 6.9$  &         ...       &         ...        &  1 \\
DES\,J022248.57$-$521651.5  &   35.70236  & $ -52.28099 $ &  22.25  &  21.81  &  21.57  &  59471.0  &   G600  &    5.9  &   \phs$57.4 \pm 9.3$  &         ...       &         ...        &  1 \\
       4745740262792352128  &   35.70283  & $ -52.28476 $ &  19.41  &  18.68  &  18.31  &  58882.1  &  G1200  &   23.7  &   \phs$53.1 \pm 1.1$  &  $1.46 \pm 0.51$  &  $-3.31 \pm 0.37$  &  1 \\
                            &   35.70283  & $ -52.28476 $ &  19.41  &  18.68  &  18.31  &  59201.2  &  G1200  &   16.4  &   \phs$52.9 \pm 1.2$  &  $1.56 \pm 0.50$  &  $-3.24 \pm 0.34$  &  1 \\
                            &   35.70283  & $ -52.28476 $ &  19.41  &  18.68  &  18.31  &  59471.0  &   G600  &   67.7  &   \phs$48.9 \pm 3.1$  &  $1.42 \pm 0.21$  &  $-3.34 \pm 0.17$  &  1 \\
DES\,J022248.90$-$521625.9  &   35.70375  & $ -52.27386 $ &  22.30  &  21.82  &  21.60  &  59471.0  &   G600  &    6.0  &   \phs$47.4 \pm 7.0$  &         ...       &         ...        &  1 \\
\hline
       4359353356893691776  &  247.71368  & $  -0.96735 $ &  19.60  &  19.02  &  18.70  &  59384.2  &  G1200  &   14.0  &     $-403.8 \pm 1.5$  &  $2.18 \pm 0.34$  &  $-2.25 \pm 0.20$  &  1 \\
                            &  247.71368  & $  -0.96735 $ &  19.60  &  19.02  &  18.70  &  59393.2  &  G1200  &   11.6  &     $-403.1 \pm 1.7$  &  $1.44 \pm 0.59$  &  $-2.73 \pm 0.47$  &  1 \\
                            &  247.71368  & $  -0.96735 $ &  19.60  &  19.02  &  18.70  &  60030.0  &  G1200  &   19.9  &     $-402.0 \pm 1.2$  &  $1.93 \pm 0.42$  &  $-2.40 \pm 0.26$  &  1 \\
       4359353155032064256  &  247.71490  & $  -0.97875 $ &  21.02  &  20.69  &  20.61  &  60030.0  &  G1200  &    3.8  &     $-399.7 \pm 3.7$  &         ...       &         ...        &  1 \\
       4359356213048800128  &  247.71795  & $  -0.88796 $ &  20.84  &  20.51  &  20.33  &  60030.0  &  G1200  &    4.9  &     $-398.9 \pm 3.1$  &         ...       &         ...        &  1 \\
       4359353150735257472  &  247.71988  & $  -0.97731 $ &  20.36  &  19.90  &  19.68  &  59384.2  &  G1200  &    6.0  &     $-403.4 \pm 3.4$  &         ...       &         ...        &  1 \\
                            &  247.71988  & $  -0.97731 $ &  20.36  &  19.90  &  19.68  &  59393.2  &  G1200  &    4.8  &     $-404.7 \pm 3.0$  &         ...       &         ...        &  1 \\
                            &  247.71988  & $  -0.97731 $ &  20.36  &  19.90  &  19.68  &  60030.0  &  G1200  &    8.7  &     $-400.8 \pm 2.2$  &         ...       &         ...        &  1 \\
       4359353155032064640  &  247.72368  & $  -0.98238 $ &  20.55  &  20.16  &  20.03  &  59384.2  &  G1200  &    4.6  &     $-403.5 \pm 4.3$  &         ...       &         ...        &  1 \\
                            &  247.72368  & $  -0.98238 $ &  20.55  &  20.16  &  20.03  &  60030.0  &  G1200  &    6.3  &     $-397.7 \pm 3.9$  &         ...       &         ...        &  1 \\
       4359353253820330624  &  247.72520  & $  -0.96842 $ &  20.72  &  20.30  &  20.12  &  59384.2  &  G1200  &    4.3  &     $-404.2 \pm 4.2$  &         ...       &         ...        &  1 \\
                            &  247.72520  & $  -0.96842 $ &  20.72  &  20.30  &  20.12  &  60030.0  &  G1200  &    6.1  &     $-405.6 \pm 2.8$  &         ...       &         ...        &  1 \\
       4359353258111283328  &  247.72604  & $  -0.96706 $ &  16.70  &  15.97  &  15.63  &  59384.2  &  G1200  &   96.3  &     $-403.2 \pm 1.0$  &  $2.53 \pm 0.20$  &  $-2.69 \pm 0.11$  &  1 \\
                            &  247.72604  & $  -0.96706 $ &  16.70  &  15.97  &  15.63  &  59393.2  &  G1200  &   84.8  &     $-403.0 \pm 1.0$  &  $2.28 \pm 0.21$  &  $-2.80 \pm 0.12$  &  1 \\
                            &  247.72604  & $  -0.96706 $ &  16.70  &  15.97  &  15.63  &  59751.1  &  G1200  &   30.1  &     $-401.9 \pm 1.1$  &  $1.90 \pm 0.32$  &  $-3.00 \pm 0.19$  &  1 \\
                            &  247.72604  & $  -0.96706 $ &  16.70  &  15.97  &  15.63  &  60030.0  &  G1200  &  126.7  &     $-403.1 \pm 1.0$  &  $2.20 \pm 0.21$  &  $-2.84 \pm 0.12$  &  1 \\
       4359354976098213248  &  247.74093  & $  -0.94068 $ &  21.13  &  20.79  &  20.61  &  60030.0  &  G1200  &    3.9  &     $-406.7 \pm 4.1$  &         ...       &         ...        &  1 \\
\enddata
\tablenotetext{a}{Source\_id for stars in the Gaia~DR3 catalog, and DES identifier for those that are not.}
\tablenotetext{b}{Quoted magnitudes represent the weighted-average PSF magnitude derived from the DES images using SourceExtractor \citep{drlica15}.}
\tablenotetext{c}{Heliocentric velocities are listed for stars and approximate redshifts are given for background galaxies.}
\tablenotetext{d}{A value of 1 indicates that the star is a member of the relevant satellite, whereas 0 indicates a non-member.}
\tablecomments{Only member stars are listed here.  The full table will be available in the online version of the journal.}
\end{deluxetable*}

%% file: abundtab.tex
\begin{deluxetable*}{lccrrrrr}
\tablecolumns{8}
\tabletypesize{\footnotesize}
\tablecaption{\label{tab:abunds}Stellar Abundances}
\tablehead{El. & $N$ & ul & $\log \epsilon$ & [X/H] & $\sigma_{\text{[X/H]}}$ & [X/Fe] & $\sigma_{\text{[X/Fe]}}$}
\startdata
\cutinhead{DELVE~1-S1}
C (C-H)   &   2 &     &$+7.25$&$-1.18$&  0.39 &$+1.63$&  0.30 \\
N (C-N)   &   2 &     &$+7.08$&$-0.75$&  0.45 &$+2.07$&  0.36 \\
O I   &   4 &     &$+7.53$&$-1.16$&  0.17 &$+1.66$&  0.26 \\
Na I  &   2 &     &$+3.98$&$-2.26$&  0.28 &$+0.55$&  0.21 \\
Mg I  &   4 &     &$+5.15$&$-2.45$&  0.14 &$+0.36$&  0.10 \\
Al I  &   1 &     &$+3.73$&$-2.72$&  0.55 &$+0.10$&  0.48 \\
Si I  &   2 &     &$+5.17$&$-2.34$&  0.34 &$+0.47$&  0.26 \\
K I   &   1 &     &$+2.80$&$-2.23$&  0.20 &$+0.58$&  0.17 \\
Ca I  &   8 &     &$+3.90$&$-2.44$&  0.12 &$+0.38$&  0.07 \\
Sc II &   4 &     &$+0.34$&$-2.81$&  0.18 &$+0.00$&  0.25 \\
Ti I  &   3 &     &$+2.22$&$-2.73$&  0.19 &$+0.08$&  0.12 \\
Ti II &  14 &     &$+2.14$&$-2.81$&  0.13 &$-0.00$&  0.15 \\
Cr I  &   4 &     &$+2.74$&$-2.90$&  0.18 &$-0.09$&  0.10 \\
Cr II &   1 &     &$+2.76$&$-2.88$&  0.20 &$-0.06$&  0.25 \\
Fe I  &  70 &     &$+4.69$&$-2.81$&  0.14 &$+0.00$&  0.02 \\
Fe II &   8 &     &$+4.71$&$-2.79$&  0.12 &$+0.02$&  0.18 \\
Co I  &   3 &     &$+2.13$&$-2.86$&  0.15 &$-0.05$&  0.14 \\
Ni I  &   1 &     &$+2.90$&$-3.32$&  0.24 &$-0.51$&  0.19 \\
Sr II &   2 &     &$-1.57$&$-4.44$&  0.25 &$-1.62$&  0.26 \\
Ba II &   4 &     &$-2.00$&$-4.18$&  0.15 &$-1.37$&  0.15 \\
Eu II &   1 & $<$ &$-1.78$&$-2.30$&\nodata&$+0.49$&\nodata\\
\cutinhead{Eri~III-S1}
C (C-H)   &   2 &     &$+7.13$&$-1.30$&  0.42 &$+1.78$&  0.27 \\
N (C-N)   &   1 &     &$+7.31$&$-0.52$&  0.67 &$+2.57$&  0.53 \\
O I   &   2 &     &$+7.86$&$-0.83$&  0.21 &$+2.25$&  0.30 \\
Na I  &   2 &     &$+3.99$&$-2.25$&  0.43 &$+0.83$&  0.30 \\
Mg I  &   5 &     &$+5.54$&$-2.06$&  0.21 &$+1.02$&  0.14 \\
Al I  &   1 &     &$+3.56$&$-2.89$&  1.31 &$+0.19$&  1.24 \\
Si I  &   1 &     &$+4.87$&$-2.64$&  0.63 &$+0.44$&  0.59 \\
K I   &   2 &     &$+2.94$&$-2.09$&  0.27 &$+1.00$&  0.20 \\
Ca I  &   4 &     &$+3.40$&$-2.94$&  0.20 &$+0.15$&  0.14 \\
Sc II &   3 &     &$+0.31$&$-2.84$&  0.29 &$+0.25$&  0.40 \\
Ti II &   4 &     &$+2.02$&$-2.93$&  0.20 &$+0.15$&  0.21 \\
Cr I  &   1 &     &$+2.84$&$-2.80$&  0.41 &$+0.29$&  0.33 \\
Fe I  &  38 &     &$+4.42$&$-3.08$&  0.25 &$+0.00$&  0.05 \\
Fe II &   2 &     &$+4.39$&$-3.11$&  0.18 &$-0.02$&  0.28 \\
Ni I  &   1 &     &$+2.91$&$-3.31$&  0.39 &$-0.23$&  0.30 \\
Sr II &   1 &     &$-2.11$&$-4.98$&  0.64 &$-1.89$&  0.60 \\
Ba II &   3 &     &$-1.70$&$-3.88$&  0.23 &$-0.79$&  0.20 \\
Eu II &   1 & $<$ &$-0.61$&$-1.13$&\nodata&$+1.97$&\nodata\\
\enddata
\end{deluxetable*}

%% file: linetab.tex
\begin{deluxetable*}{lrrrrrrrrrrrrrrrrr}
\tablecolumns{17}
\tabletypesize{\footnotesize}
\tablecaption{\label{tab:lines}Line Measurements}
\tablehead{Star & $\lambda$ & ID & $\chi$ & $\log gf$ & EW & $\sigma$(EW) & FWHM & ul & $\log \epsilon_i$ & $\sigma_i$ & $e_i$ & $s_X$ & $\delta_{i,\teff}$ & $\delta_{i,\logg}$ & $\delta_{i,\nu_t}$ & $\delta_{i,\text{[M/H]}}$ }
\startdata
DELVE~1-S1 & 6300.30 &   8.0 &  0.00 & -9.69 &  15.8 &  3.2 &  0.23 &   &  7.30 &  0.20 &  0.09 &  0.10 & +0.10 & +0.10 & -0.00 & +0.01 \\
DELVE~1-S1 & 5528.40 &  12.0 &  4.35 & -0.55 &  52.9 &  3.6 &  0.23 &   &  5.15 &  0.14 &  0.06 &  0.10 & +0.08 & -0.02 & -0.02 & +0.00 \\
DELVE~1-S1 & 4077.71 &  38.1 &  0.00 &  0.15 &  syn  &  syn &  0.15 &   & -1.64 &  0.28 &  0.10 &  0.10 & -0.09 & +0.20 & -0.10 & -0.01 \\
DELVE~1-S1 & 4313.00 & 106.0 &\nodata&\nodata&  syn  &  syn &  0.14 &   &  7.29 &  0.34 &  0.01 &  0.10 & +0.31 & -0.10 & +0.00 & +0.03 \\
DELVE~1-S1 & 4129.70 &  63.1 &  0.00 &  0.22 &  syn  &  syn &\nodata&$<$& -1.78 &\nodata&\nodata&\nodata&\nodata&\nodata&\nodata&\nodata \\
Eri~III-S1 & 6300.30 &   8.0 &  0.00 & -9.69 &  29.2 &  5.1 &  0.23 &   &  7.76 &  0.22 &  0.09 &  0.10 & +0.16 & +0.08 & -0.01 & +0.02 \\
Eri~III-S1 & 5528.40 &  12.0 &  4.35 & -0.55 &  77.5 &  6.5 &  0.24 &   &  5.41 &  0.20 &  0.09 &  0.10 & +0.14 & -0.03 & -0.04 & +0.00 \\
Eri~III-S1 & 4077.71 &  38.1 &  0.00 &  0.15 &  syn  &  syn &  0.14 &   & -2.11 &  0.60 &  0.54 &  0.10 & +0.23 & +0.04 & -0.04 & +0.04 \\
Eri~III-S1 & 4313.00 & 106.0 &\nodata&\nodata&  syn  &  syn &  0.22 &   &  7.17 &  0.37 &  0.06 &  0.10 & +0.35 & -0.00 & -0.00 & +0.02 \\
Eri~III-S1 & 4129.70 &  63.1 &  0.00 &  0.22 &  syn  &  syn &\nodata&$<$& -0.61 &\nodata&\nodata&\nodata&\nodata&\nodata&\nodata&\nodata \\
\enddata
\tablecomments{A portion of this table is shown for form. The full
  version is available online. The columns are wavelength ($\lambda$), species (ID), excitation potential ($\chi$), cross section ($\log gf$), equivalent width and uncertainty (EW and $\sigma$(EW)), full width half max (FWHM), upper limit flag (ul), line abundance and total uncertainty ($\log \epsilon_i$, $\sigma_i$), statistical uncertainty ($e_i$), per-line systematic uncertainty $s_X$, and errors due to stellar parameter uncertainties ($\delta_{i,\teff}$, $\delta_{i,\logg}$, $\delta_{i,\nu_t}$, $\delta_{i,\text{[M/H]}}$). The total uncertainty $\sigma_i$ is the quadrature sum of $e_i$, $s_X$, $\delta_{i,\teff}$, $\delta_{i,\logg}$, $\delta_{i,\nu_t}$, and $\delta_{i,\text{[M/H]}}$.}
\end{deluxetable*}
